\documentclass[aps,physrev,twocolumn,superscriptaddress,floatfix]{revtex4-2}
\usepackage{graphicx}
\usepackage{dcolumn}
\usepackage{bm}
\usepackage{comment}
\usepackage{color}
\usepackage{siunitx}
\usepackage{ulem}
\usepackage{placeins}
\usepackage{amsmath}
\usepackage{hyperref}
\hypersetup{breaklinks=true}

\begin{document}

\title{Single cells can resolve graded stimuli}

\author{Mirna Kramar}
 \affiliation{Institut Curie, Physics of Cells and Cancer, UMR 168, 75005 Paris, France}

\author{Lauritz Hahn}
\affiliation{Laboratoire de physique de l'\'Ecole normale sup\'erieure, 75005 Paris, France}

\author{Aleksandra M Walczak*}
\affiliation{Laboratoire de physique de l'\'Ecole normale sup\'erieure, CNRS, PSL University, Sorbonne Universit\'e, and Universit\'e de Paris, 75005 Paris, France}
\thanks{Contact authors: \\ aleksandra.walczak@ens.fr \\ thierry.mora@ens.fr \\ mathieu.coppey@curie.fr \\}

\author{Thierry Mora*}
\affiliation{Laboratoire de physique de l'\'Ecole normale sup\'erieure, CNRS, PSL University, Sorbonne Universit\'e, and Universit\'e de Paris, 75005 Paris, France}
\thanks{Corresponding author e-mail: thierry.mora@ens.fr}

\author{Mathieu Coppey*}
 \affiliation{Institut Curie, Physics of Cells and Cancer, UMR 168, 75005 Paris, France}
 \thanks{Corresponding author e-mail: mathieu.coppey@curie.fr}%

\date{\today}

\begin{abstract}
Cells use signalling pathways as windows into the environment to gather information, transduce it into their interior, and use it to drive behaviours.
MAPK (ERK) is a highly conserved signalling pathway in eukaryotes, directing multiple fundamental cellular behaviours such as proliferation, migration, and differentiation, making it of few central hubs in the signalling circuitry of cells. 
Despite this versatility of behaviors, population-level measurements have reported low information content (\textless 1 bit) relayed through the ERK pathway, rendering the population barely able to distinguish the presence or absence of stimuli. Here, we contrast the information transmitted by a single cell and a population of cells. Using a combination of optogenetic experiments, data analysis based on information theory framework, and numerical simulations we quantify the amount of information transduced from the receptor to ERK, from responses to singular, brief and sparse input pulses. We show that single cells are indeed able to resolve between graded stimuli, yielding over 2 bit of information, however showing a large population heterogeneity.
\end{abstract}


\maketitle

\section{\label{introduction}Introduction}

After decades of almost exclusive focus on behaviour, learning and decision-making in neural organisms, the attention is slowly shifting not only to non-neural organisms, but also the building blocks of all organisms: single cells~\cite{Gershman2021}.
Information processing is at the heart of these phenomena. Understanding how the information is taken up, relayed and integrated is crucial to decoding the way decisions are made and behaviour is executed on all levels of life.

In cells, information about the environment is relayed through signalling pathways. The signal taken up at the receptor is transduced through a complex network of proteins which act as biochemical modification cascades (e.g. by phosphorylation or conformational change), propagating the signal towards the nucleus for regulation of gene transcription, or to other hubs of the signalling circuitry, ultimately leading to behavioural decisions such as proliferation, migration, or apoptosis.

The existence of a range of cellular decisions resulting from a comparatively small number of signalling hubs inspired a number of studies aimed at elucidating cellular information processing strategies~\cite{Perkins2009}. In particular, the bulk of these studies focused on the mitogen-activated protein kinase (MAPK) pathway, a signalling pathway implicated in key cellular functions. The most intensely studied sub-network of the MAPK pathway -- commonly referred to as the ERK signalling pathway -- is a cascade of four kinases: Ras, Raf, MEK and ERK~\cite{Huang1996}.

A deep dive into the intricacies of ERK pathway signalling reveals a multitude of isoforms, mechanisms and feedbacks~\cite{Ram2023, Ram2023a}. One of the essential questions about information processing in a signalling cascade is its sensitivity to signals, which can be addressed from different perspectives -- by asking questions about the nature of ligands, binding affinity, ligand dynamics, concentrations, and noise~\cite{Potter2017,Grabowski2019, Cheong2011, Vazquez-Jimenez2019}. In the context of concentration, the ability to resolve between graded inputs can be quantified as the amount of information relayed when a stimulus is applied at different magnitudes. Traditionally, signalling pathways have been regarded as binary switches -- active or inactive, devoid of the ability to resolve between different input states. Within the framework of information theory, the amount of information is estimated in bits: 1 bit for an on/off switch, and \textgreater 1 bit for the ability to resolve graded inputs~\cite{Bialek2012,Bauer2021}. 

The studies addressing the information content in key signalling pathways reported a surprisingly small information content. A study of ERK and NFkB signalling pathways showed that the information capacity is less than 1 bit when considering a single feature of the response signal, such as the amplitude of the response, albeit around 1.5 bit when taking into account dynamical features of the response~\cite{Selimkhanov2014}. However, the input stimulus in the form of a diffusible ligand remains present after initial application, resulting in a continuous and adaptive response. Another recent study presented a higher information content (2.3 bit), however attributed it to multimodal perception (combinatorial encoding through several features including cell-intrinsic and environmental properties), stressing that non-contextual information content in the ERK pathway does not exceed 0.7 bit~\cite{Kramer2022}. Effectively, this means that by relying solely on information transduced through the ERK pathway, cells are unable to distinguish between the presence and absence of stimuli, i.e. falling short of operating as an on/off switch.

Given the essential role of the ERK pathway in directing a range of core cellular functions, the reported low resolution between inputs is surprising and suggests that the question of the amount of information carried by the ERK pathway is still open~\cite{Suderman2018}. Namely, the methods used in the listed studies present several challenges which could have a significant impact on the measured information quantity. First, by averaging out the responses to the stimulus across the cell population, the individuality of the cells arising from the differences in their biochemical content is erased~\cite{Elowitz2002,Levchenko2014}. If the cell-to-cell variability in the response to stimuli (i.e. the extrinsic noise) is high, averaging the responses over the population could render the amount of information low for an average cell, regardless of a possible high information content in single cells~\cite{Mitchell2018,Tay2010}. Next, single-cell quantification within the information-theoretic framework requires building probability distributions, i.e. recording responses of single cells to repeated stimulus. This is not reliably achievable with diffusible ligands due to the effects of ligand trapping, and changing cell medium may impact cellular functions, introducing additional uncertainty~\cite{Vis2020}. Finally, relying on diffusible ligands allows only a small number of input states. Given that the maximum theoretical amount of information that can be transduced is given by the number of input states, having a range of well-resolved input states is essential~\cite{Levchenko2014}. A recent study reporting higher information content in the ERK pathway highlights the importance of single-cell measurements~\cite{Goetz2024}, however in the absence of single-cell data, it relies on theoretical modelling for quantification. An earlier experimental work suggested that the information content based solely on the amplitude of the response may exceed 1 bit when accounting for cell-to-cell variability~\cite{Toettcher2014}.

Here, we revisit the question of information content in the ERK signalling pathway by introducing an experimental and quantification strategy to overcome the challenges faced by the studies presented above. We perform optogenetic stimulations of the MAPK pathway, carrying out an unprecedented number of repeated stimulations with seven light doses, and quantify the mutual information in single cells using an algorithm specifically developed for application in cell signalling~\cite{Jetka2019}. We find a large variation in the encoding features of single cells within the population, demonstrating the crucial importance of single-cell quantification. We show that some cells transmit over 2 bit of information, close to the maximal possible value set by the 7 input stimuli (2.8 bits) which is significantly higher than previously reported for single pulses (i.e. non-sustained) of input. This result confirms our intuition about the cell's ability to non-contextually resolve graded inputs through the ERK pathway. Furthermore, to account for the limitations of our setup, we perform numerical simulations of optogenetic activation of ERK. The results of the simulations corroborate our experimental findings and highlight the importance of repeated stimulation. Finally, we look for patterns behind the intrinsic differences in information content in the population, and discuss their implications in the functioning of the cell collective, as well as the generation of behaviour.

\begin{figure*}[h!]
    \includegraphics[width=\linewidth]{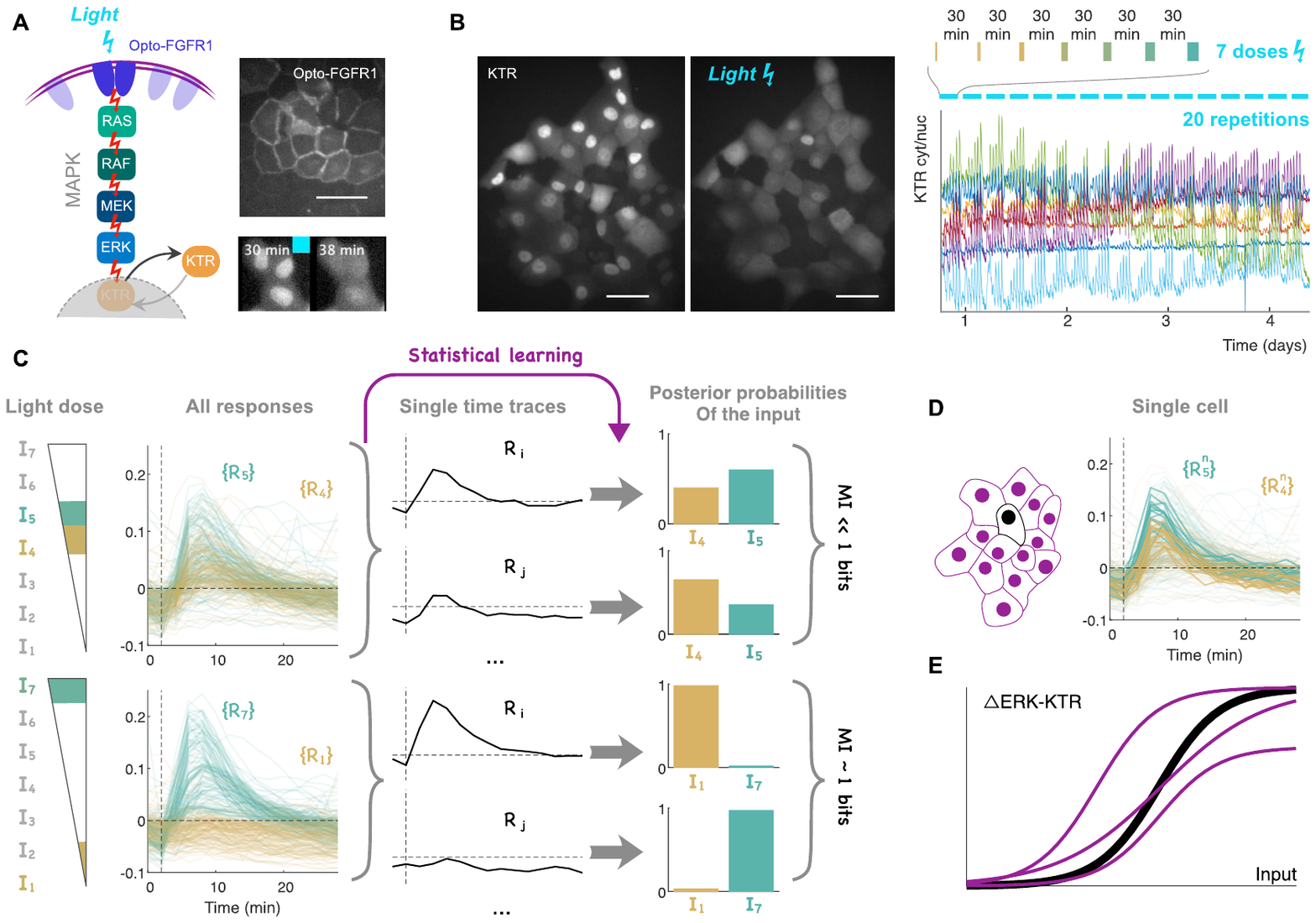} 
       \caption{\textbf{Measuring mutual information of the MAPK pathway with many repeats at the single cell level}.\\ \textbf{(A)} Scheme of the MAPK signalling cascade activated by light. The optogenetic actuator OptoFGFR1\cite{Kim2014} consists of the fusion between truncated fibroblast growth factor receptor 1 (cytosolic domain, without the extracellular part) and CRY2, together with a fluorescent reporter (mCitrine). The construct localizes at the plasma membrane (top fluorescent image, scale bar is 100µm), and when illuminated with blue light CRY2 oligomerizes leading to the transphophorylation of the cytosolic tail of FGFR1 and thus receptor activation. Once activated, the receptor activates the MAP kinase phosphorylation cascade ending in the double phosphorylation of ERK, ppERK. We used ERK-KTR-mKate2 as a reporter of ERK activity, which is a minimal substrate domain that shuttles out of the nucleus when phosphorylated by ppERK (bottom fluorescent images, before activation and 8 minutes after, scale bar is 100µm). \textbf{(B)} Round of optogenetic stimulations. Left, example of the whole field of view during the acquisition, before activation and 5 minutes after a light pulse (full time series available in~\textit{Supplemental Material} as Video 1 and Video 2). Right, ratios of cytosolic over nuclear ERK-KTR intensities for 7 selected cells over the whole duration of the stimulation routine that consists in 20 repetitions of 7 light pulses of increasing dose 30 minutes apart. \textbf{(C)} Estimation of mutual information (MI). The 7 light doses of increasing strength lead to a gradual increase of cells' ERK responses. For two close light doses (here the 4\textsuperscript{th} and the 5\textsuperscript{th} light doses, $I_4$ and $I_5$), there is a strong overlap of the compiled cell responses (all selected cells and all repeats, shown in the top plot as $\{R_4\}$ -gold- and $\{R_5\}$ -turquoise-); whereas for the two extreme light doses ($I_1$ and $I_7$) cell responses appear distinguishable. In our pipeline, the statistical learning-based SLEMI algorithm~\cite{Jetka2019} assigns posterior probabilities of the input stimulation. For clarity of presentation, here we only show the posterior probabilities of the input for two inputs. When cell responses have a large overlap, the posterior probabilities of the input are close to 0.5 (top, low mutual information), whereas when cell responses are distinguishable the probabilities are close to 1 (bottom, mutual information close to its maximum of 1 bit for two states). \textbf{(D)} Single cell responses are more reproducible than the whole population. When only one cell is selected among the field of view (black cell), even for two close stimuli the cell responses appear more distinguishable than the whole population (10 repeats for $I_4$ and $I_5$ stimuli of the best performing cell are represented over all cells and repeats responses). \textbf{(E)} Each cell responds specifically. The differences of response overlap between all cells and selected cell can be explained by the fact that every cell has its own perception of stimuli. Here, the dose-response curve of neighboring (purple) cells could diverge from the selected (black) cell by a shift in the half-maximum, a change of sensitivity (slope), or dynamic range (amplitude).}
       \label{fig:optointro}
\end{figure*}

\section*{{\label{results}Results}}
\subsection*{\label{results:optosetup}Optogenetics allows for accurate estimation of mutual information in the ERK signalling pathway}

There has been a number of constraints impeding the quantification of information flow in the ERK signalling pathway using the information-theoretic framework: limited number of well-resolved inputs, small number of samples due to inability to perform repeated single-cell stimulation, limitations in the biosensor dynamic range, noise in the measurement, and non-optimal distribution of stimuli with respect to the sensitivity of the pathway.

We tackle these constraints by building an experimental setup and a data analysis routine presented in Fig.~\ref{fig:optointro}. We mimic the varying input concentrations on the Fibroblast growth factor receptor (FGFR) by using its optogenetic analogue, OptoFGFR1~\cite{Kim2014} stably introduced in MCF10A cells, and apply single short pulses of blue laser light of different strengths to activate the pathway. It was previously shown that this optogenetic molecular system responds gradually to light doses \cite{Dessauges2022}, and we optimize the light doses to span the whole dynamic range of cell responses (\textit{Supplemental Material}, Fig. S1). 

The kinase translocation reporter (KTR) is a fast-reacting and ERK1/2-specific fluorescent reporter~\cite{Delacova2017} which reversibly shuttles out of the nucleus when ERK is activated (Fig.~\ref{fig:optointro}-A). Using a fluorescent histone label H2B-iRFP for nuclear segmentation, we quantify the ratio between cytosolic and nuclear KTR, thus obtaining a ratiometric readout of ERK activation robust to fluctuations in laser intensity and focus.

We carry out simultaneous stimulation of the ERK pathway in each cell in the population using 7 light doses of increasing intensity, then repeat the cycle 20 times in one dataset, and 15 times in the other dataset presented here, with 30 min between each stimulus to allow for ERK deactivation, the total imaging time being 93.6 and 81.6 hours, respectively. Additionally, in the main dataset (20 repeats), we measure the light intensity of each light dose with a power meter mounted above the imaging dish in the light path of the laser. This simultaneous recording of the light power allows for quantification of variability at the level of the input, estimated to be around 10 to 20\% (\textit{Supplemental Material}, Fig. S2 and S3.)

Figure~\ref{fig:optointro}-B shows the raw trajectories of ERK activation in individual cells over the course of the experiment. We obtained 20 cells across four different stage positions in the main dataset (20 repeats), and 40 cells in the other dataset (15 repeats) presented here. We applied strict criteria for subsequent analysis: only cells which display a reliable, continuous signal and stable morphology over the whole course of the experiment were considered. Such long imaging times come with significant caveats: cells divide, migrate out of the field of view, undergo changes in morphology due to interactions within the population, all of which can render their signal non-continuous or unreliable for further quantification. Passing of fluorescent debris floating over the imaged population equally presents challenges for signal extraction. 

It is also worthwhile to note that the cells are imaged in the presence of serum, which is an important distinction from the usual protocol for imaging ERK activity used in existing studies. Although the presence of the serum enhances basal ERK activity~\cite{Aoki2013}, it is necessary for cell survival on the timescale of days, and supports physiological conditions, which is desirable when discussing the results in the context of organismal function.

The quantification pipeline is presented in Fig.~\ref{fig:optointro}-C and detailed in~\textit{Materials and methods}, Appendix~\ref{matmethods}. The time series of single-cell ERK activity over the duration of the whole experiment are detrended by a moving averaging window and subsequently fragmented into 30 min trajectories, starting with the application of the stimulus and ending one timepoint before the application of the next stimulus. To account for the fact that cell responses are time varying signals that may contain information beyond a single given timepoint, we extracted a vector containing the values of the response over a temporal window. Likewise, we also applied a dimensionality reduction method (principal component analysis, PCA) to compress the whole response trajectories into few components. 

Next, we used these features (vector containing ERK responses and principal components) as input in the Statistical Learning Estimator of Mutual Information (SLEMI)~\cite{Jetka2019}, which we found to be best-performing among several mutual information estimators we tested according to different criteria (\textit{Appendix}~\ref{appendix-MIestimation} and \textit{Supplemental Material} Fig.~S4). The algorithm estimates information-theoretic quantities between a finite state input and a continuous output using logistic regression~\cite{Jetka2019}.

For each input condition $x_i$, $i=1,...,N$, the experiment is repeated to obtain $n_i$ samples $y_i^j$, $j=1,...,n_i$. SLEMI uses multinomial logistic regression to infer the distribution $P(x_i|y_i^j)$ from the data, such that the MI can be approximated as 
\begin{equation}
I(X;Y)\approx \sum_{i=1}^N \frac{P(x_i)}{n_i}\sum_{j=1}^{n_i}\log_2\left(\frac{P(x_i|y_i^j)}{P(x_i)}\right).
\end{equation}

The importance of performing the analysis on single cell data is outlined in Fig.~\ref{fig:optointro}-D,E. The varying biomolecular makeup of each cell necessitates single-cell analysis since the individuality of each given cell may distort their perceptions of stimulus intensity. For the estimation of mutual information, this implies that only repeated measurements of the response of a single cell will overcome the extrinsic variability, which manifests as a lack of reproducibility at the population level.

\begin{figure*}[th!]
\includegraphics[width=0.8\linewidth]{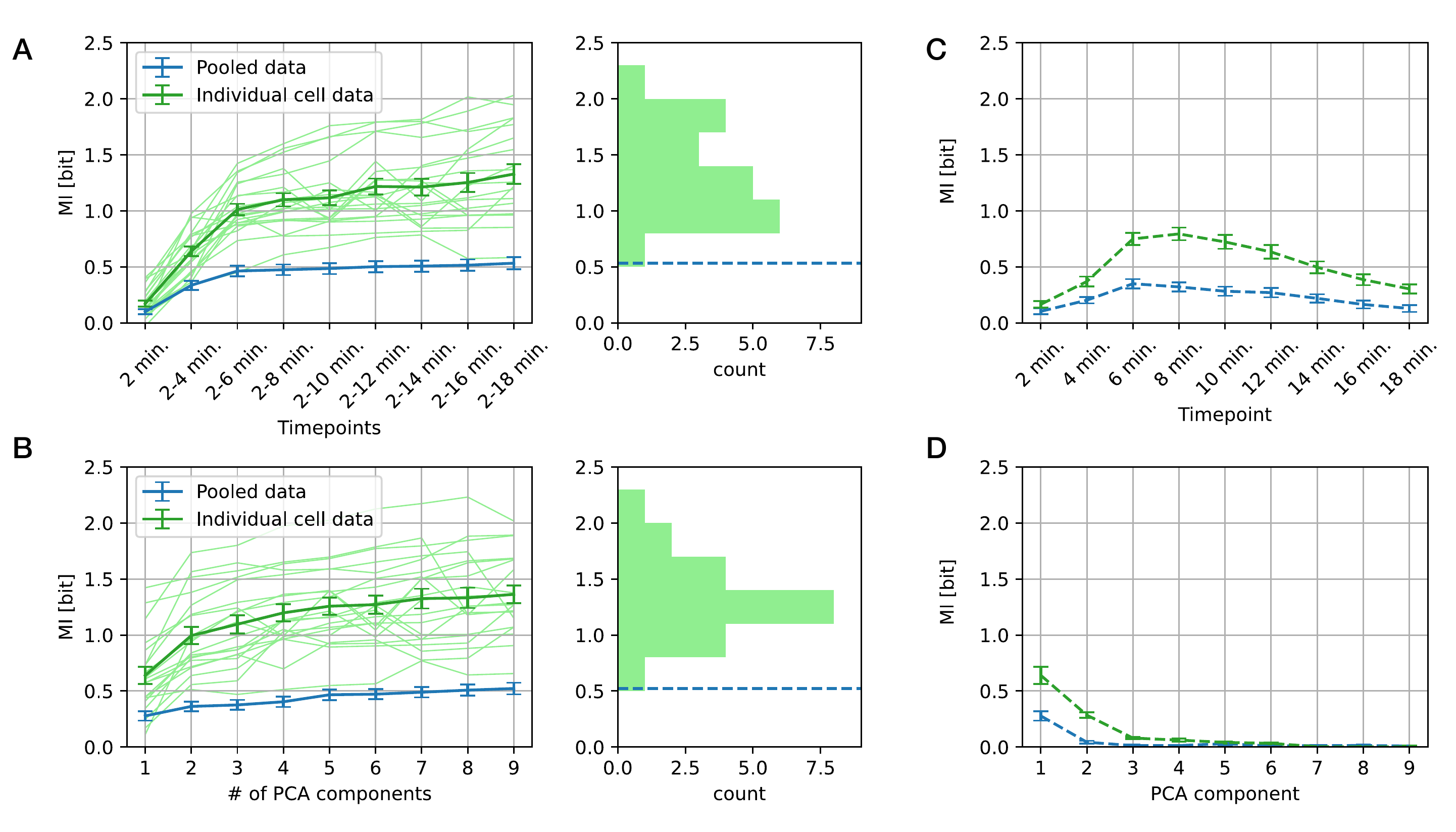}
\caption{\textbf{Population-level readout erases the input resolution abilities of single cells. \\
(A)} Mutual information contained in single cells vs. the population. Mutual information calculated by taking single-cell trajectories (light green) and by pooling the trajectories and thereby erasing the cell identities (blue) shown for increasing number of timepoints, sampled every two minutes. Average information quantity contained in single-cell trajectories is shown in dark green. The histogram depicts the distribution of the mutual information extracted from the last data point of single-cell trajectories (green), the dashed blue line marking while the mutual information value for pooled trajectories.~\textbf{(B)} Analogously, quantification of mutual information for an increasing number of principal components (sorted by variance) of response trajectories, and the corresponding histogram of mutual information values.~\textbf{(C)} Mutual information from single-cell trajectories contained in individual timepoints (given in minutes after light stimulus) for the average of the single-cell data (green) and pooled trajectories (blue). ~\textbf{(D)} Analogously, the mutual information contained in individual principal components for single-cell trajectories (green) and pooled trajectories (blue).}
\label{fig:main-result}
\end{figure*}

\subsection*{Population-level readout erases the input resolution abilities of single cells}
We quantify the mutual information (MI) between the values of the light doses (input) and the time trajectories of the cytoplasmic-to-nuclear ratio of ERK-KTR fluorescence (output). The results of the quantification are depicted in Fig.~\ref{fig:main-result}. These temporal trajectories were compressed by selecting timepoints or by PCA, as described in the section above. In the first case, we initially only consider the first timepoint after stimulation, then successively calculate the MI using trajectories over time windows of increasing length. In the second case, we start with only the first PCA component and then add further components in order of decreasing variance.

Given that we used 7 light doses, the maximum MI is set by $\log_2(7) \approx$ 2.8 bit. When considering individual cells, the population average of the mutual information reaches a plateau around 1.5 bit when 4 timepoints or 4 principal components (or more) are taken. However, when all cell responses are pooled, the maximal information content drops to $\approx$ 0.5 bit, a low value showing that the pooled responses of the cells are strongly overlapping and preventing the stimuli to be distinguished. This mutual information can be slightly increased to $\approx$ 0.7 bit considering only the lowest and highest stimuli (\textit{Supplemental Material} Fig.~S5), with the population still unable to clearly distinguish between the presence and absence of stimuli.

The higher mutual information in single cells compared to the population experimentally confirms the crucial importance of single-cell analysis, and shows that single cells are able to resolve between several intensities of the stimulus. Furthermore, the results reveal a large variability in information content within the population (Fig.~\ref{fig:main-result}), with some cell responses yielding over 2 bit of information, the highest being 2.2 bit. Such a large value shows that cells can almost reach the maximal theoretical mutual information.
 
\subsection*{Information is contained in response dynamics}
We next investigate the informative abilities of individual timepoints in the cell responses, as well as in individual principal components obtained by dimensionality reduction (PCA). We find that the third and fourth timepoint are the most informative with around 1 bit of information (Fig.~\ref{fig:main-result}). The same amount of information is obtained from the first PCA component. The third and fourth timepoints are closest to the peak of the response trajectories, making it the most informative feature of the response. By reading out just the response peak, single cells can on average distinguish between presence and absence of the stimulus. When considering pooled cell responses, the information content of single timepoints follows a similar trend as for single cell responses, but without the ability to discriminate between stimuli.

\subsection*{Cell-to-cell variability in responses is large but appears to be unrelated to endogenous ERK fluctuations}

Having observed large differences in information content, we delve deeper into cell-to-cell variability to understand how information content degrades when pooling cell responses. We first introduce a signal decoder to compare the true input with the input that is predicted by the cell. The predicted input is the dose with the highest assigned probability in the multinomial logistic regression performed in the SLEMI algorithm, averaged over the different repeats of the stimulus. Thus, the predicted dose index for input dose $i$ is given by $(1/n_i)\sum_j\text{argmax}_k P(x_k|y_i^j)$, with $n_i$ and $P(x_k|y_i^j)$ as described above. 

\begin{figure*}[tbh!]
\includegraphics[width=0.6 \linewidth]{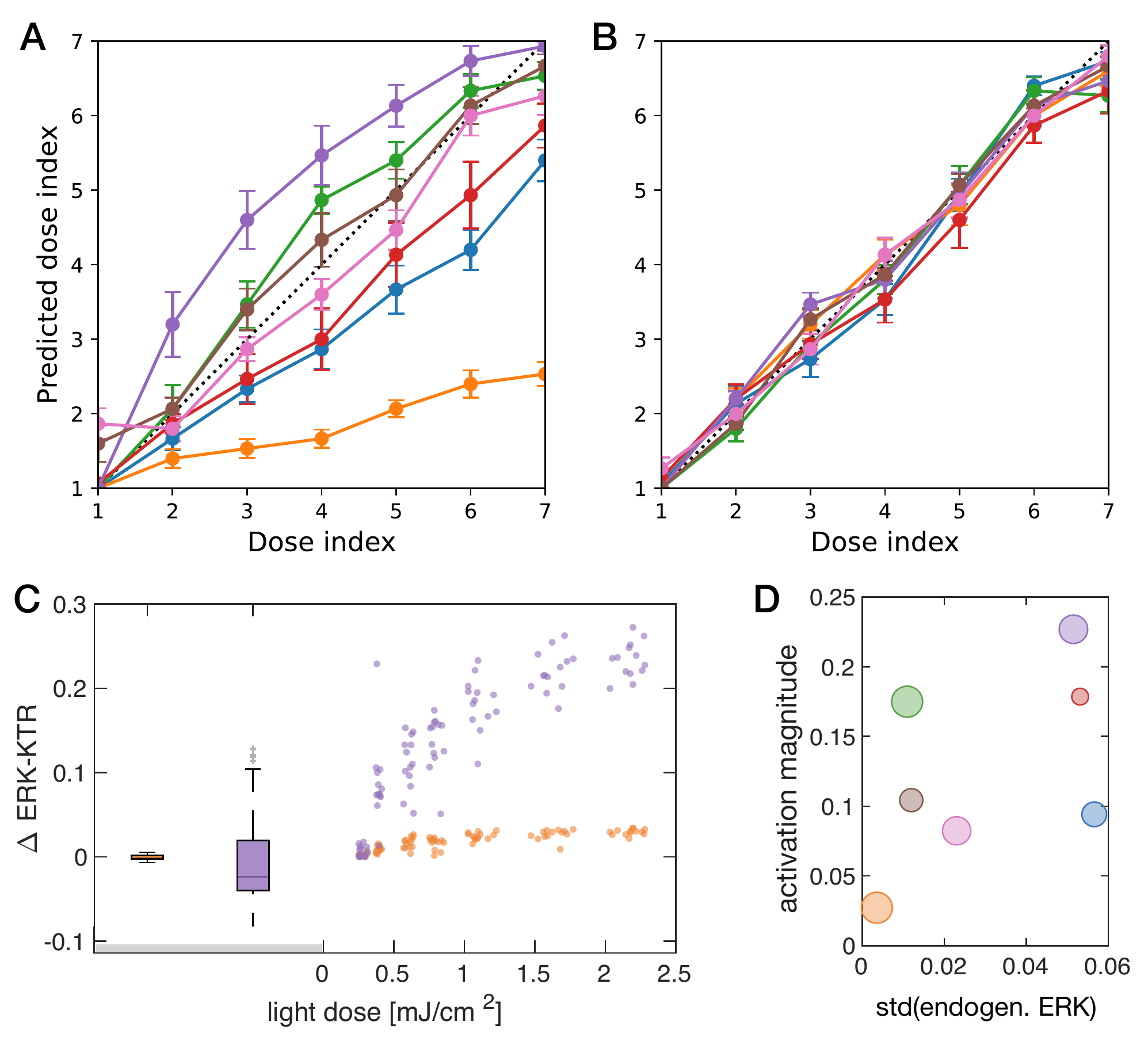}
\caption{\textbf{Large intrinsic variability demonstrates the need for individual decoders. (A,B)} Light dose inferred by the cell against the actual light dose using the pooled cell model decoder (A) and individual cell model decoder (B), calculated for all cells in one stage position. Unbiased decoding is shown as a dotted black line. \textbf{(C)} Dose-response relationships of the two most extreme cells from the panels (A-B) above, here as scatter plots shown together with the spontaneous fluctuations in ERK activity (as bar plot) in the absence of stimuli (marked by gray bar). \textbf{(D)} Using the same color code, the maximal response magnitude of the cells, i.e. the difference between $\Delta$ERK-KTR for the highest and lowest light dose, is shown against the standard deviation of the endogenous fluctuations. The mutual information content for each cell is depicted by the size of the marker. The association between the mutual information measured in cells, the magnitude of ERK activation, and the magnitude of endogenous fluctuations was tested using Spearman's Rank Correlation and Kendall’s Tau test, which did not report a statistically significant correlation.}
\label{fig:decoder}
\end{figure*}

Using the decoder based on the pooled cell model, we compare the decoding abilities of individual cells belonging to a population at one stage position, i.e. cells which received the same input doses. These results are presented in Fig.~\ref{fig:decoder}-A, where unbiased decoding is depicted by a dotted black line across the diagonal. While none of the cells display unbiased decoding using the pooled cell model, some cells perform close to population average, while other cells strongly diverge from it. The changing slope for individual cells reveals an interesting feature: the cells' decoding behaviour changes with the light doses. This behaviour is unlikely to be time-dependent since the cell model is calculated from repeats occurring periodically over the course of the whole experiment (days), and more likely informs about a cell's sensitivity to a dose range with respect to the population. The need for individual (single-cell) decoders is shown in~Fig.~\ref{fig:decoder}-B, where all cells perform close to unbiased decoding of the input signal (dotted black line).

Next, we focus on the potential role of endogenous fluctuations in ERK activity in response to stimuli and information encoding. High endogenous fluctuations, coming from either spontaneous activities or cell-to-cell communication, may sum up with the optogenetic activation and deteriorate the amount of information being transduced. In Fig.~\ref{fig:decoder}-C, we compare the magnitude of endogenous ERK activity fluctuations during the imaging phase without stimulation to the magnitude of ERK response to light stimuli. As an example, we compare two cells whose decoding abilities deviated most from the prediction using the pooled cell model in Fig.~\ref{fig:decoder}-A. The cell with high endogenous fluctuations (in purple) shows a high ERK response magnitude, and the cell with low endogenous fluctuations (in orange) displays comparatively low ERK response magnitude, hinting that the endogenous ERK activity could alter the response to stimuli. However, not all cells exhibit this behaviour (\textit{Supplemental Material}, Fig.~S6).

To further delve into this question, we compare the mutual information of each of the seven cells while simultaneously depicting the maximum ERK response to stimuli against the standard deviation of endogenous ERK fluctuations (Fig.~\ref{fig:decoder}-D). Due to the small sample size, we use both Spearman's Rank Correlation and Kendall’s Tau test to assess the relationships between these three variables. First, we find a weak positive correlation between endogenous fluctuations and maximum ERK response, however not statistically significant in either test (Spearman’s $\rho$ = 0.393, p = 0.396 Kendall’s $\tau$ = 0.238, p = 0.562). Next, endogenous fluctuations and mutual information show a moderate negative correlation (Spearman’s $\rho$ = -0.643, p = 0.139, Kendall’s $\tau$ = -0.429, p = 0.239), suggesting that higher endogenous fluctuations could be associated with lower mutual information. However, this association was not statistically significant, although it did approach significance in Spearman’s test. Finally, the correlation between response to stimuli and mutual information is very weak and negative, and not statistically significant in either test (Spearman’s $\rho$ = -0.179, p = 0.713, Kendall’s $\tau$ = -0.238, p = 0.562).

\begin{figure*}[tbh!]
\includegraphics[width= 0.6 \linewidth]{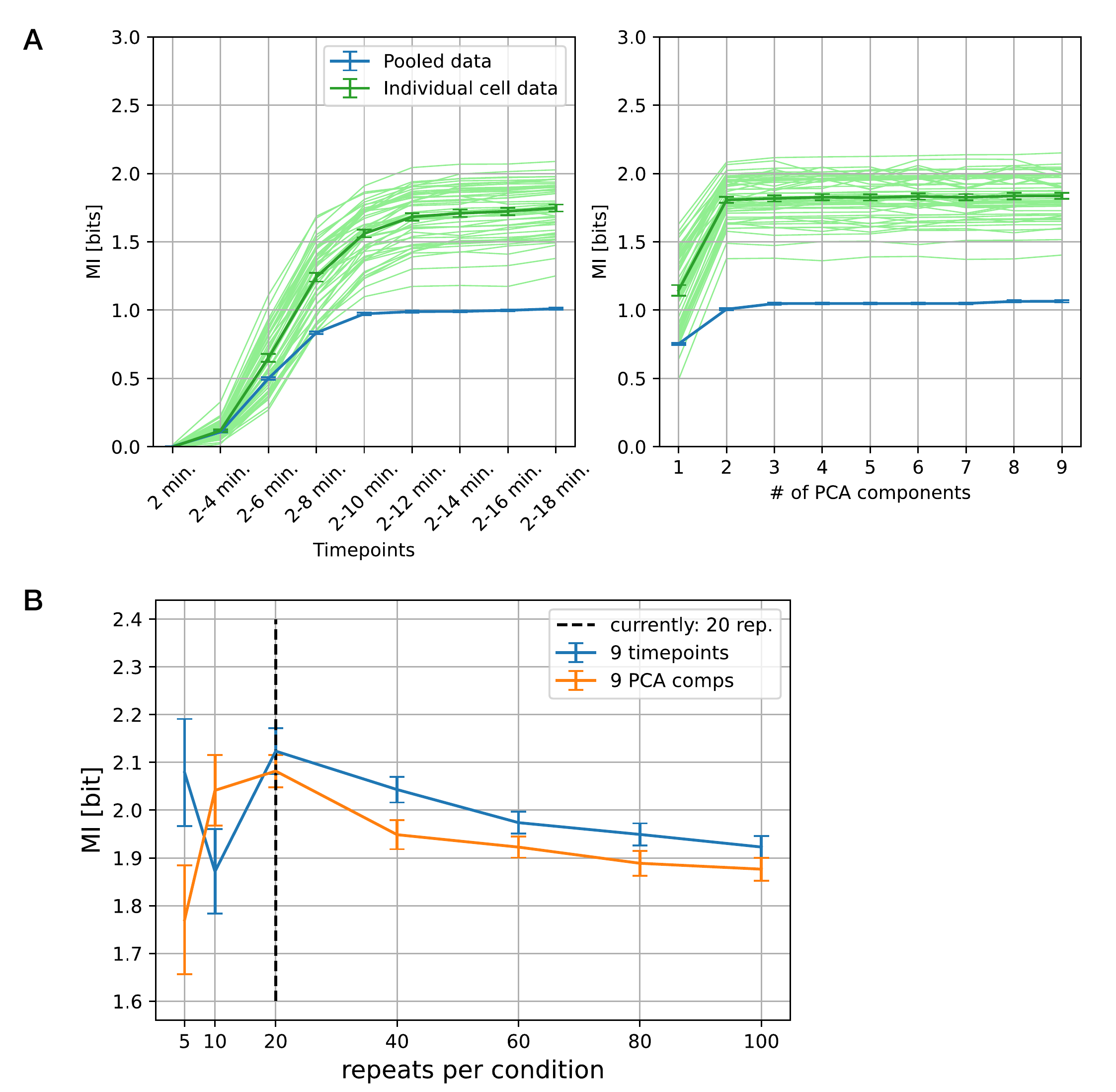}
\caption{\textbf{Numerical simulations corroborate experimental findings. (A)} Simulations of optogenetic activation of the MAPK cascade, akin to the experiments. Analogously to Fig.~\ref{fig:main-result}, mutual information contained in the simulated cells' ERK response trajectories was calculated using increasing number of timepoints (left) and PCA (right). The results from pooled cell responses is shown in blue, and the results from single cell responses in green (individual cells in light green, the average of single cell responses in dark green). For the clarity of presentation, a subset of 40 simulated cells was selected for this visualisation. \textbf{(B)}~The estimation of mutual information in simulated trajectories here for different number of repeats per condition. The information is estimated using the first 9 timepoints (blue) and 9 PCA components (orange). The value of mutual information undergoes fluctuations for low number of repeats, and stabilizes after $\approx$ 100 repeats. The black dashed line is drawn at 20 repeats per condition, the current experimental limit. }
\label{fig:simulations}
\end{figure*}

Given the absence of convincing evidence of association, the endogenous ERK activity did not prove to be a driving factor causing cell-to-cell variability, and optogenetic stimulations might overcome these spontaneous activities irrespectively of their strengths. Taken together, we conclude that the main driver of cell-to-cell variability is a variable encoding scheme that is specific for each cell.

\subsection*{Numerical model corroborates experimental results and bridges scales}

Having a sufficient number of samples is a known challenge for accurate estimation of information within the information-theoretic framework~\cite{Ish-Horowicz2017,Hernandez2022}. Even with the unprecedented number (20) of samples, i.e. repeated stimulation of single cells that we present here, inferring the reliability of the information estimate remains a challenge in this complex biological system. To this end, we use a published kinetic model of the ERK pathway with parameters sampled from experimental data~\cite{Dessauges2022}, and perform numerical simulations of optogenetic activation of the pathway with light doses mimicking the experimental ones (\textit{Supplemental Material} Fig.~S7). We create a set of 400 virtual cells by generating random combinations of parameters, allowing a variation of up to 18\% for each parameter to account for cell-to-cell variability, and perform 1000 repeats of stimulation with each light dose. The resulting simulation trajectories are further modified by adding $0.01$ Gaussian white noise to match experimental results.

Then, we subject the obtained response trajectories to the same analysis pipeline as used for the experimental data. Figure~\ref{fig:simulations}-A depicts the information contained in the time traces using the two feature reduction techniques already described above. The results of the simulations agree with the experimental findings: the information contained in single-cell responses greatly surpasses the one contained in pooled data. The average single cell in the simulations performs better than its experimental counterpart likely due to a lesser degree of additive noise, which in the experiments arises from various sources: biological, experimental, and post-processing. The mutual information contained both in the pooled and single-cell trajectories rapidly increases with decreasing noise added to the trajectories (\textit{Supplemental Material}, Fig.~S8), however, it is important to note that even with very low noise, the information content in best-performing simulated cells barely exceeds 2.5 bit. This information content, smaller than the theoretical maximum (2.8 bit), indicates the upper bound due to the variability of the input.

\begin{figure*}[tb!]
\includegraphics[width=0.7\linewidth]{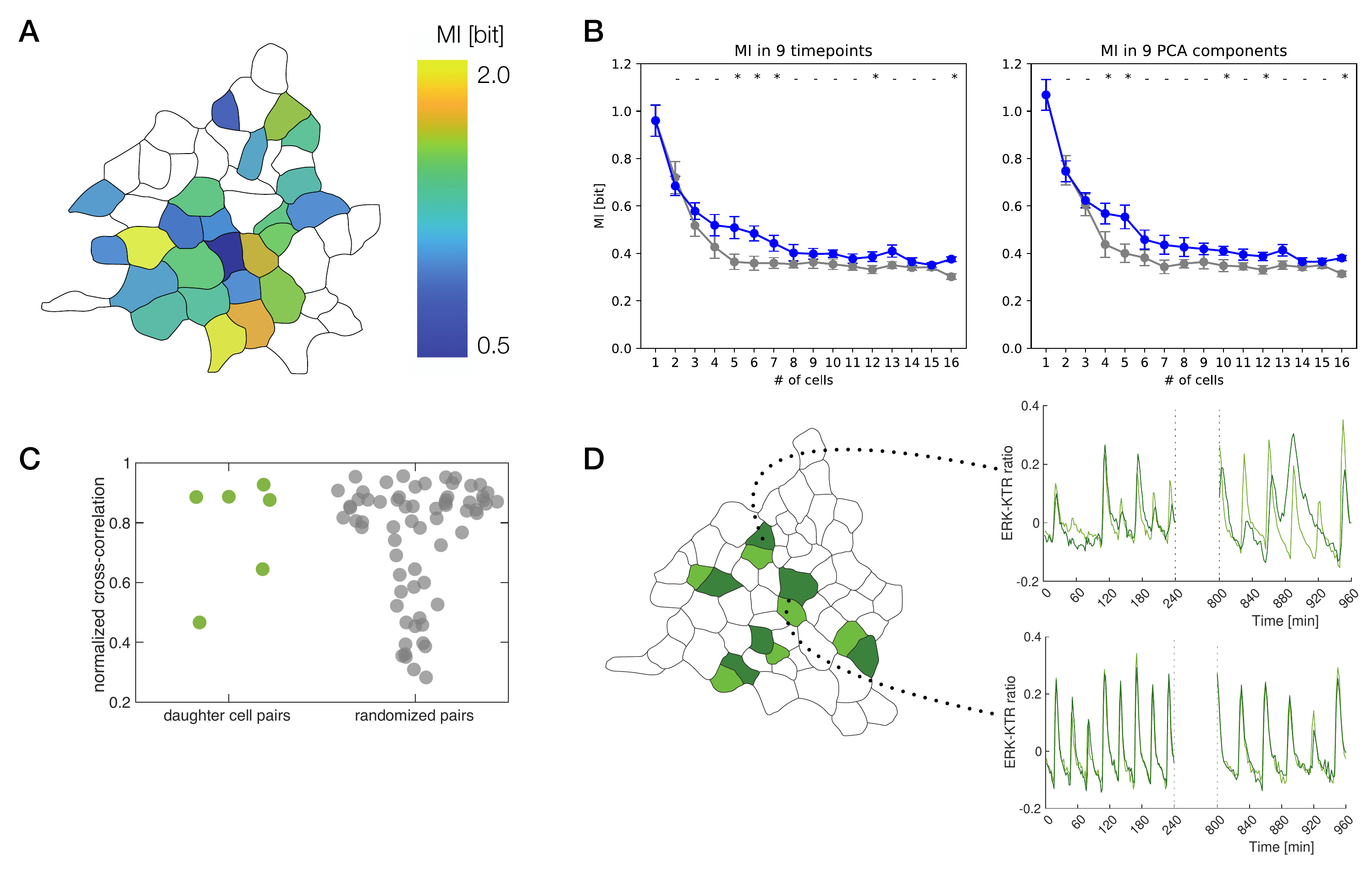}
\caption{\textbf{Spatio-temporal aspects of Mutual information in the cell collective.}\\ 
\textbf{(A)} Spatial distribution of mutual information in a cell population, obtained from 9 timepoints of the response trajectories (left). Only the cells which passed selection criteria for analysis are taken into account, otherwise shown in white.~\textbf{(B)} Mutual information (from 9 timepoints and PCA 9 components, as described above) calculated for clusters of neighbouring cells of different sizes (number of cells in a cluster on x axis) compared to the mutual information in clusters of the same size, but with cells drawn from random positions in the population. The curves depicting the information in clusters of actual neighbours falls off slightly slower compared to the randomized clusters. Top of the plot: Mann-Whitney U test, with (-) and (*) indicating a p-value above 0.05 and below 0.05, respectively.
\textbf{(C)} Comparison of the values of cross-correlation of ERK response signal between two daughter cells to randomized pairs shows no indication of increased correlation in the pair. Tested by Kolmogorov-Smirnov test, $p=0.3744$. \textbf{(D)} Example ERK response trajectories for the marked daughter cell pairs, where two extreme cases are shown: uncorrelated cell responses (top) and highly correlated ones (bottom).}
\label{fig:neighbours-and-daughters}
\end{figure*}

The satisfying degree of resemblance between the simulated and experimental cells now allows for probing the role of the number of samples, i.e. stimulation repeats, on the estimate of mutual information (Fig.~\ref{fig:simulations}-B). Both time-point based trajectories and PCA components used as input result in strong fluctuations in MI value for small sample sizes. After approximately 100 repeats, the curve stabilises around a lower value of about 1.9 bit. The initial fluctuations of the MI value indicate that the number of repeats in our experiment (marked with a dashed line) is the minimum required for a reliable estimate, being at the start of a monotonic behaviour of the curve. This result shows that the experimental values of mutual information should be taken with some caution: simulations show that we might be overestimating the mutual information for our experimental sample size, however only by about 10 percent. In the case of low noise levels, the MI reaches a stable value of about 2.5 bit already for 20 samples (\textit{Supplemental Material}, Fig.~S8). These results further emphasise the role of noise in the estimation of mutual information.

\subsection*{Spatio-temporal aspects of information in a cell collective}
We further address the question of cell-to-cell variability by examining potential spatial patterns of information in a cell collective. The experimental findings presented above show that mutual information drops strongly from one cell to the whole population (Fig.~\ref{fig:main-result}-A). Here, two scenarios can be envisaged. First, neighbouring cells could present a higher degree of similarity, either because they originated from the same cell, or due to their spatial proximity, whereby in both cases the mutual information would slowly decay when considering increasing groups of close-by cells. Alternatively, cells may have different identities, irrespective from their spatial distribution within the population, and mutual information would drop as the group of cells increases due to their variability in encoding inputs. To address this question, we use the other dataset (15 repeats) containing a higher number of cells per stage position which passed the set criteria for analysis and also form a large patch of cells. Figure~\ref{fig:neighbours-and-daughters}-A depicts the spatial distribution of mutual information in the population of cells, with no obvious spatial pattern. A comparison of the mutual information in a patch containing a different number of neighbouring cells with the information in a randomized patch shows that mutual information presents a slight spatial component that can be observed for groups of 4-7 cells. However, neither smaller nor bigger groups present any statistically significant increase in mutual information compared to randomized groups, indicating a lack of a convincing spatial pattern.

To further test if two neighboring cells are as variable as the whole population, we focused on the most extreme case of anticipated similarity: daughter cells, i.e. two cells resulting from the same division event. Over the course of the experiment, six division events took place in this cell population. Since they appear at different times and therefore do not satisfy the selection criteria for MI quantification, we cannot estimate the mutual information for these cells. However, we can examine the similarity of daughter cells' responses to light stimuli, and hence address the question of cell-to-cell variability. Figure~\ref{fig:neighbours-and-daughters}-B depicts the results of cross-correlation of the ERK response signal for pairs or daughter cells and randomized pairs. We used MATLAB built-in two-sample Kolmogorov-Smirnov test with the null hypothesis stating that both samples come from the same distribution. The test failed to reject the the null hypothesis at the 5\% significance level ($p=0.3744$), meaning that in line with our previous observation, there is no significant difference in correlation between the two groups. This result indicates that daughter cells are as variable as two cells randomly picked from the population. 

\section*{Discussion}

In this work, we addressed the question of information content in the ERK signalling pathway, with the specific aim of clearing the long-standing conundrum in the field which arose from experimental obstacles. We designed an experimental setup to carry out repeated optogenetic stimulations of single cells with inputs of seven different strengths, resulting in a large number of response trajectories. We evaluated the trajectories for information content using the information theory formalism, and obtained mutual information measured in bits.

The experimental results show that pooling the ERK response trajectories from all cells results in a large loss of information which renders the population barely able to distinguish between the presence and absence of stimuli, while single-cell response trajectories contain over 1.5 bit of information on average, with a fraction of cells yielding over 2 bit of information, i.e. effectively being able to distinguish between more than 4 input states. This result showcases the crucial importance of single-cell analysis, and explains the apparent low information content reported previously. 

Our numerical simulations of the optogenetic ERK pathway activation support the experimental results, as well as provide intuition about the accuracy of the mutual information estimate for a varying number of samples. We believe that our estimation of the mutual information can be taken as a lower bound given the experimental limitations. The measurements are bounded by experimental noise, including the biosensor dynamics, image acquisition and data analysis routines which are always imperfect, as well as the variability of the input itself (here 10 to 20 \%), and of course of the number of different stimuli.

Another possible factor contributing to noise is the slowly evolving trend in ERK responses (\textit{Supplemental Material} Fig.~S9). Since the imaging takes over 4 days, the cells may drift in their own encoding scheme. Some cells display a monotonous drift over time (\textit{Supplemental Material} Fig.~S10). Thus, the mutual information might be higher if the estimation could be performed on a shorter time window. 

We delved deeper to identify the sources of the observed cell-to-cell variability, finding no evidence of the variability being linked to endogenous ERK fluctuations, and no convincing indication of a spatial pattern. Interestingly, we found that the variability from cell to cell was due to the differences in their encoding schemes, i.e. the cell-specific dose-response curves. This cellular specificity seems not to be inherited, since the responses of daughter cells were not more correlated than randomized pairs. It could be hypothesized that cells are equipped with adaptation mechanisms to either maintain their sensitivity set-point, or to filter out this extrinsic variability in the downstream integration~\cite{Wilson2017, Tkacik2008} of ERK signals. 

Our results agree with the previous findings of information being contained in response dynamics~\cite{Selimkhanov2014}, which show that single signalling pathways can encode more than an on/off state. As introduced earlier, cells can also distribute information through a combinatorial activation of several pathways~\cite{Kramer2022}. We would like to stress here that the two mechanisms are not mutually exclusive, and an increase of information in a given pathway will further enhance the ability of cells to encode more subtle cues at the level of the whole signalling circuitry.

It is surprising that the magnitude of endogenous fluctuations of ERK activity does not seem to have a bearing on the cell's ability to encode information. In our experiment, the cells are imaged in physiological conditions, in the presence of serum, which normally activates the ERK pathway thanks to the presence of a variety of growth factors. However, we think that endogenous ERK fluctuations should not be hastily dismissed. Intriguingly, in the absence of optogenetic stimulation we observed significant endogenous fluctuations on the same scale as the responses to the first two weakest inputs. Yet, those weak optogenetic inputs were eliciting reproducible and distinguishable cellular responses. Thus, cells may have a fast adaptation mechanisms that allow signalling events to 'shortcut' other ones. Alternatively, the repeated stimulation routine may have entrained the cells to lower their basal activity on a relatively long timescale ($\approx$1hr). It is interesting to notice that spontaneous ERK activity is happening in the lower range of response dynamics. In the context of our information-centered approach, it seems that cells do not exploit their full information capacity, or that we are underestimating the amount of mutual information in the lower range of inputs. It would be interesting for future studies to quantitatively assess the prior distribution of inputs in the generation of behaviour, for example during development~\cite{Delacova2017}. 

The observed endogenous fluctuations may also reflect cell-to-cell communication, since a cell's ERK activity is communicated to the neighbours~\cite{Aoki2013}. However, the absence of correlation between daughter cells' ERK responses and the absence of clear spatial patterns suggest that cells respond to our stimuli in an autonomous fashion. The direct experimental assessment of the role of cell-to-cell communication would require the use of isolated cells, but unfortunately, due to the very long duration of the experiment and the tendency of the cell line to grow in clusters, recording ERK responses to repeated stimuli on isolated cells is not achievable.

The nature of the input is another important aspect to consider when quantifying information in signalling pathways. A number of recent studies demonstrated the key role of input dynamics on cellular behaviour~\cite{Stanoev2020,Aikin2020, Gagliardi2021, Ender2022}, and it is likely that temporally modulated signals may encode more information or higher specificity in cell response. It will be paramount for the future studies to assess the quantitative aspects of the inputs in physiological conditions -- whether the most significant part of the input comes from neighboring cells or from the ever-changing broader environment.

Likewise, in the pioneering study on the topic of information in signalling pathways~\cite{Selimkhanov2014}, the input (ligand concentration) was sustained throughout the measurement, possibly allowing a reinforcement of the response and resulting in a high information content on the population level. In contrast, our study examined the response of the cells to very short pulses, spaced out in time to allow the return of ERK activity to its basal level. This important distinction accounts for a different biological scenario, where cells experience sparse pulses of stimulation (e.g. during apoptosis of neighbouring cells~\cite{Gagliardi2021}). The surprising variability in information content within the population invites questions about the heterogeneity in sensitivity to inputs, and its implications on subsequent cellular decision-making. 

Overall, with the introduction of repeated stimulation and single-cell level quantification, this work removed a previously daunting obstacle to understanding cellular information processing, and opened a way to studying the link between information encoding and organization of behaviour in populations of cells.

\begin{acknowledgments}
This project has received funding from the European Union’s Horizon 2020 research and innovation programme under the Marie Skłodowska-Curie grant agreement No.101031499.
The experimental setups and infrastructure used were supported by the LabEx Cell(n)Scale (ANR-10-LABX-0038), Labex and Equipex IPGG (ANR-10-NANO0207), Idex Paris Science et Lettres (ANR-10-IDEX-0001-02 PSL), French National Research Infrastructure France-BioImaging (ANR-10-INBS-04), Institut Convergences Q-life (ANR-17-CONV-0005). We kindly thank Carine Vias for the preparation of plasmids for lentiviral transduction, Olivier Pertz for the MCF10A cell line and discussions, as well as Jan Mikelson and Mustafa Kammash for sharing their model and parameters of the MAPK pathway.
\end{acknowledgments}

\appendix

\section{Mutual Information Estimation}
\label{appendix-MIestimation}
Several algorithms exist in the literature to estimate the mutual information between random variables, for example k-nearest-neighbour-based (knn) methods \cite{Kraskov2004,Selimkhanov2014}, Markov model-based methods \cite{Tang2021}, and methods using Bayesian inference \cite{Jetka2019}. We have tested and benchmarked several of these algorithms (\textit{Supplemental Material} Fig.~S4), and found that the Bayesian inference-based SLEMI~\cite{Jetka2019} algorithm performed best with our data.

For an input $X$ taking $N$ different, discrete values $x_i$ and the continuous, time-dependent output $Y(t)$, the mutual information between $X$ and $Y(t)$, in bit, reads
\begin{align}
&I(X;Y)= \\&\sum_{i=1}^N P(x_i)\int \mathcal D[y(t)] P(y(t)|x_i)\log_2\left(\frac{P(x_i|y(t))}{P(x_i)}\right), \nonumber
\end{align}
where $\int\mathcal D[y(t)]$ is a path integral over all possible trajectories $y(t)$. When repeating the experiment for a given input $x_i$ a total of $n_i$ times, the distribution $P(y(t)|x_i)$ is sampled. The SLEMI algorithm \cite{Jetka2019} uses the law of large numbers to approximate the MI as
\begin{equation}
I(X;Y)\approx \sum_{i=1}^N \frac{P(x_i)}{n_i}\sum_{j=1}^{n_i}\log_2\left(\frac{P(x_i|y_i^j)}{P(x_i)}\right),
\end{equation}
where $y_i^j$, $j=1,...,n_i$ are the experimentally sampled trajectories. SLEMI then uses statistical learning to infer $P(x_i|y_i^j)$ from the data. Following the authors, we use multinomial logistic regression for the inference step.

The other algorithms mentioned above estimate $P(y_i^j|x_i)$ to compute mutual information. However, this is a high-dimensional distribution and is generally difficult to estimate, so in general, SLEMI requires fewer samples per input condition to converge. In our experiments, only a very limited number of samples can be obtained from each cell ($\sim 20$ per cell), so SLEMI's faster convergence is a significant advantage. Furthermore, SLEMI is a parameter-free estimator: unlike knn estimators, where an arbitrary value of $k$ must be fixed, SLEMI does not contain any numerical parameters that need to be fixed.

Finally, the available knn-based estimators each have disadvantages relevant to the problem at hand. In particular, the original estimator by Kraskov, Stögbauer and Grassberger \cite{Kraskov2004} does not use discrete, but rather continuous variables as inputs, and we previously found that the algorithm by Selimkhanov \emph{et al.} \cite{Selimkhanov2014} struggles with high-dimensional outputs \cite{Hahn2023}.

Nonetheless, care must be taken to avoid overfitting. When using 7 different input conditions and a 9-dimensional output (9 timepoints or PCA components), $(7-1)*9=54$ parameters need to be fitted. To avoid overfitting, we found it necessary to add 5-fold cross-validation and a L1 regularization penalty, to which end we re-implemented SLEMI in Python. Since the L1 norm is not invariant under linear transformations, we expect slight variations in the model depending on the parametrization of the data.

\section{Materials and methods}
\label{matmethods}
\subsection*{Cell culture}
\label{cellculture}

The MCF10A breast epithelium cell line stably expressing the OptoFGFR-mCitrine was kindly gifted by the lab of Olivier Pertz (Universität Bern, Institute of Cell Biology). Subsequently, lentiviral transduction was used to generate a stable cell line expressing the H2B-iRFP histone marker, and the ERK-KTR-mKate2 reporter of nuclear ERK activity.
The cells were cultured in plastic-bottom cell culture flasks and standard MCF10A medium. The medium contains \SI{500}{mL} DMEM F-12 (Sigma), GlutaMAX (1\%, Gibco), Horse serum (5\%, Sigma), Human EGF (\SI{20}{ng/mL}, Sigma), Human insulin (0.1\%, Sigma), Cholera toxin (\SI{100}{ng/mL}, Sigma), Hydroxycortisone (\SI{0.5}{mg/mL}, Sigma), and Penicillin-Streptomycin (1\%, ThermoFisher Scientific). 

\subsection*{Imaging setup}
\label{imagingsetup}
The imaging was performed on an epifluorescence imaging setup, composed of an Olympus IX83 inverted microscope, electron-multiplying Evolve EMCCD camera (Photometrics), and a laser bench (Gataca Systems, Massy, France) equipped with the following lasers used here: 491nm (activation of OptoFGFR1), 560nm (excitaton of ERK-KTR-mKate2 reporter), and 640nm (excitation of the H2B nuclear marker). The cells were imaged using a 20x Olympus air objective (NA 0.7).

The atmosphere in the stage top incubator is controlled by the CO2-Controller 2000 (Pecon) and the temperature controller TempControl 37-2 digital (Pecon).
The laser illumination is controlled by the iLas2 software (Gataca Systems, Massy, France), and the imaging setup is executed using Metamorph software (\textit{Molecular Devices}, Eugene, OR) , through a custom-made Matlab script. 

The light power of each light dose used for OptoFGF stimulation was measured with a ThorLabs PM100D power meter, and the measurement was executed automatically by a custom-made Matlab script. The sensor of the power meter was fixed to the glass cover of the stage-top incubator and protected from ambient light. 

\subsection*{Live cell imaging and optogenetic activation}
\label{imaging}

The cells were plated \SI{24}{h}-\SI{36}{h} before the experiment on 35-mm glass-bottom FluoroDish (Thermo Fisher), 120 000 cells per dish in \SI{2}{mL} FluoroBrite DMEM (Thermo Fisher) medium with 5\% Horse Serum (Sigma), 1\% GlutaMAX (Gibco), and 1\% Penicillin-Streptomycin (Sigma). The addition of serum ensures cell viability and normal physiology over the multi-day course of the experiment. 

The acquisition was carried out every \SI{2}{min}, and images were acquired in two channels. Images of the H2B-iRFP nuclear marker were acquired by~\SI{640}{nm} laser excitation for \SI{200}{ms}, and the images of the ERK-KTR-mKate2 reporter were acquired by~\SI{640}{nm} laser excitation for \SI{200}{ms}. 

The excitation of the OptoFGFR optogenetic actuator with the 488 laser was performed every \SI{30}{min} during the stimulation phase, starting \SI{120}{min} from the start of imaging for the main dataset with 20 repeats, and \SI{100}{min} for the other dataset with 15 repeats. After all the stimulation repeats are carried out, the imaging is continued for several hours. The phase without stimulation at the beginning and the end of imaging serves two purposes: to estimate the endogenous variations of ERK activity, and to test for possible entrainment of ERK oscillations by optogenetic stimulation.
The interval of \SI{30}{min} between two stimuli allows for the return of nuclear ERK activity to its pre-stimulation level.

The optogenetic stimulation was carried out in sets of 7 light doses, and each set was repeated 20 times. The light doses are changed by modulating the illumination time in an increasing order: \SI{200}{ms}, \SI{300}{ms}, \SI{500}{ms}, \SI{700}{ms}, \SI{1000}{ms}, \SI{1500}{ms}, and \SI{2000}{ms}, while the laser power was kept constant at about \SI{200}{\micro W} in order to diminish the effects of non-linear laser power output. Previously, a wide range of light doses was applied to obtain dose-response curves throughout several experiments with many cells to identify the maximal sensitivity range of an average cell, and the chosen light doses were chosen to fall within this range. Given the duration of the cell cycle, cell migration and other effects which set a limit to the number of stimulations that can be applied on a single cell, the pattern of stimulation with a repeating set of increasing light doses was chosen over a randomised one to maximise the chances of obtaining the highest number of repeats per light dose.

\subsection*{Data analysis}
\label{datanalaysis}
The image segmentation and tracking of the nuclei in the H2B channel were carried out in iLastik software (version 1.4.0). Next, the segmented nuclei were used to extract the average nuclear ERK reporter signal from the images in the ERK-KTR channel. Equally, the average ERK reporter signal contained in a thin ring around the nucleus was extracted for the subsequent calculation of the nuclear-to-cytosolic ratio of the reporter. Ring thickness of \SI{4}{px} provides the most robust readout and the best trade-off between acquiring enough cytoplasmic signal and remaining within the boundaries of the cell in cases of the nucleus being in close proximity to the membrane. 
Using a custom-written Matlab data analysis pipeline, the nucleus-tracking trajectories were matched with the ERK signal readout, assigning each cell a unique ID. The movement of the cells or fluorescent debris passing over the field of view causes occasional issues in segmentation and tracking, resulting in missing frames -- i.e. no information on the ERK activity. To ensure data quality, only the cells with ERK signal in 95\% of the total number of frames are selected for the next step of the analysis. This criterion automatically excludes cells which underwent mitosis, migration in or out of the field of view during the acquisition, and cells with significant morphological changes, latter of which could indicate aberrant cell physiology. 
The magnitude of nuclear ERK activity is quantified as the ratio between the cytosolic and the nuclear ERK signal described above.
To calculate the dose-response curves, single-cell ERK time trajectories were detrended using a moving averaging window of 100 frames before being divided into intervals of 15 frames, starting with the timepoint of the stimulation. The raw and detrended trajectories, as well as the impact of different detrending methods on the estimation of mutual information is presented in \textit{Supplemental Material} Fig.~S11-S13.
The first two principal components obtained by principal component analysis of the ensemble of response trajctories are depicted in \textit{Supplemental Material} Fig.~S14.
The light doses are calculated as the integral of the light power measured with the power meter over the time interval of stimulation. 

\subsection*{Numerical simulations}
\label{simulations}
Numerical simulations of optogenetic MAPK pathway activation were carried out by implementing the kinetic model from~\cite{Dessauges2022} in MATLAB. The activation of ERK is read out as the ratio between the cytosolic and nuclear KTR, analogously to the experiments. The parameters -- reaction rates and initial concentrations -- were determined from experimental data~\cite{Dessauges2022} and inferred using the Nested Sampling algorithm described in~\cite{Mikelson2020}, kindly provided by the authors.
The virtual cells were created by generating 400 sets of parameters with up to 18\% randomly applied Gaussian white noise. To mimic the experimental variability in the light dose, up to 10\% Gaussian white noise was applied to the light input parameter. The light doses used in the simulations were chosen to yield activation trajectories and dose-response curves which closely resemble experimental ones (\textit{Supplemental Material} Fig.~S7). Additionally, a saturation threshold for activation was determined based on the experiments. Each of the 400 virtual cells were stimulated with the 7 light doses, spaced in time to allow KTR to return to its initial value, repeating this process 1000 times.

\sloppy
\bibliography{bibliography.bib}

\begin{thebibliography}{38}%
\makeatletter
\providecommand \@ifxundefined [1]{%
 \@ifx{#1\undefined}
}%
\providecommand \@ifnum [1]{%
 \ifnum #1\expandafter \@firstoftwo
 \else \expandafter \@secondoftwo
 \fi
}%
\providecommand \@ifx [1]{%
 \ifx #1\expandafter \@firstoftwo
 \else \expandafter \@secondoftwo
 \fi
}%
\providecommand \natexlab [1]{#1}%
\providecommand \enquote  [1]{``#1''}%
\providecommand \bibnamefont  [1]{#1}%
\providecommand \bibfnamefont [1]{#1}%
\providecommand \citenamefont [1]{#1}%
\providecommand \href@noop [0]{\@secondoftwo}%
\providecommand \href [0]{\begingroup \@sanitize@url \@href}%
\providecommand \@href[1]{\@@startlink{#1}\@@href}%
\providecommand \@@href[1]{\endgroup#1\@@endlink}%
\providecommand \@sanitize@url [0]{\catcode `\\12\catcode `\$12\catcode `\&12\catcode `\#12\catcode `\^12\catcode `\_12\catcode `\%12\relax}%
\providecommand \@@startlink[1]{}%
\providecommand \@@endlink[0]{}%
\providecommand \url  [0]{\begingroup\@sanitize@url \@url }%
\providecommand \@url [1]{\endgroup\@href {#1}{\urlprefix }}%
\providecommand \urlprefix  [0]{URL }%
\providecommand \Eprint [0]{\href }%
\providecommand \doibase [0]{https://doi.org/}%
\providecommand \selectlanguage [0]{\@gobble}%
\providecommand \bibinfo  [0]{\@secondoftwo}%
\providecommand \bibfield  [0]{\@secondoftwo}%
\providecommand \translation [1]{[#1]}%
\providecommand \BibitemOpen [0]{}%
\providecommand \bibitemStop [0]{}%
\providecommand \bibitemNoStop [0]{.\EOS\space}%
\providecommand \EOS [0]{\spacefactor3000\relax}%
\providecommand \BibitemShut  [1]{\csname bibitem#1\endcsname}%
\let\auto@bib@innerbib\@empty
\bibitem [{\citenamefont {Gershman}\ \emph {et~al.}(2021)\citenamefont {Gershman}, \citenamefont {Balbi}, \citenamefont {Gallistel},\ and\ \citenamefont {Gunawardena}}]{Gershman2021}%
  \BibitemOpen
  \bibfield  {author} {\bibinfo {author} {\bibfnamefont {S.~J.}\ \bibnamefont {Gershman}}, \bibinfo {author} {\bibfnamefont {P.~E.}\ \bibnamefont {Balbi}}, \bibinfo {author} {\bibfnamefont {C.~R.}\ \bibnamefont {Gallistel}},\ and\ \bibinfo {author} {\bibfnamefont {J.}~\bibnamefont {Gunawardena}},\ }\href {https://doi.org/10.7554/eLife.61907} {\bibfield  {journal} {\bibinfo  {journal} {Elife}\ }\textbf {\bibinfo {volume} {10}},\ \bibinfo {pages} {1} (\bibinfo {year} {2021})}\BibitemShut {NoStop}%
\bibitem [{\citenamefont {Perkins}\ and\ \citenamefont {Swain}(2009)}]{Perkins2009}%
  \BibitemOpen
  \bibfield  {author} {\bibinfo {author} {\bibfnamefont {T.~J.}\ \bibnamefont {Perkins}}\ and\ \bibinfo {author} {\bibfnamefont {P.~S.}\ \bibnamefont {Swain}},\ }\href {https://doi.org/10.1038/msb.2009.83} {\bibfield  {journal} {\bibinfo  {journal} {Mol. Syst. Biol.}\ }\textbf {\bibinfo {volume} {5}},\ \bibinfo {pages} {1} (\bibinfo {year} {2009})}\BibitemShut {NoStop}%
\bibitem [{\citenamefont {Huang}\ and\ \citenamefont {Ferrell}(1996)}]{Huang1996}%
  \BibitemOpen
  \bibfield  {author} {\bibinfo {author} {\bibfnamefont {C.~Y.~F.}\ \bibnamefont {Huang}}\ and\ \bibinfo {author} {\bibfnamefont {J.~E.}\ \bibnamefont {Ferrell}},\ }\href {https://doi.org/10.1073/pnas.93.19.10078} {\bibfield  {journal} {\bibinfo  {journal} {Proc. Natl. Acad. Sci. U. S. A.}\ }\textbf {\bibinfo {volume} {93}},\ \bibinfo {pages} {10078} (\bibinfo {year} {1996})}\BibitemShut {NoStop}%
\bibitem [{\citenamefont {Ram}\ \emph {et~al.}(2023{\natexlab{a}})\citenamefont {Ram}, \citenamefont {Murphy}, \citenamefont {DeCuzzi}, \citenamefont {Patankar}, \citenamefont {Hu}, \citenamefont {Pargett},\ and\ \citenamefont {Albeck}}]{Ram2023}%
  \BibitemOpen
  \bibfield  {author} {\bibinfo {author} {\bibfnamefont {A.}~\bibnamefont {Ram}}, \bibinfo {author} {\bibfnamefont {D.}~\bibnamefont {Murphy}}, \bibinfo {author} {\bibfnamefont {N.}~\bibnamefont {DeCuzzi}}, \bibinfo {author} {\bibfnamefont {M.}~\bibnamefont {Patankar}}, \bibinfo {author} {\bibfnamefont {J.}~\bibnamefont {Hu}}, \bibinfo {author} {\bibfnamefont {M.}~\bibnamefont {Pargett}},\ and\ \bibinfo {author} {\bibfnamefont {J.~G.}\ \bibnamefont {Albeck}},\ }\href {https://doi.org/10.1042/BCJ20230276} {\bibfield  {journal} {\bibinfo  {journal} {Biochem. J.}\ }\textbf {\bibinfo {volume} {148}},\ \bibinfo {pages} {887} (\bibinfo {year} {2023}{\natexlab{a}})}\BibitemShut {NoStop}%
\bibitem [{\citenamefont {Ram}\ \emph {et~al.}(2023{\natexlab{b}})\citenamefont {Ram}, \citenamefont {Murphy}, \citenamefont {DeCuzzi}, \citenamefont {Patankar}, \citenamefont {Hu}, \citenamefont {Pargett},\ and\ \citenamefont {Albeck}}]{Ram2023a}%
  \BibitemOpen
  \bibfield  {author} {\bibinfo {author} {\bibfnamefont {A.}~\bibnamefont {Ram}}, \bibinfo {author} {\bibfnamefont {D.}~\bibnamefont {Murphy}}, \bibinfo {author} {\bibfnamefont {N.}~\bibnamefont {DeCuzzi}}, \bibinfo {author} {\bibfnamefont {M.}~\bibnamefont {Patankar}}, \bibinfo {author} {\bibfnamefont {J.}~\bibnamefont {Hu}}, \bibinfo {author} {\bibfnamefont {M.}~\bibnamefont {Pargett}},\ and\ \bibinfo {author} {\bibfnamefont {J.~G.}\ \bibnamefont {Albeck}},\ }\href {https://doi.org/10.1042/BCJ20230277} {\bibfield  {journal} {\bibinfo  {journal} {Biochem. J.}\ }\textbf {\bibinfo {volume} {480}},\ \bibinfo {pages} {1909} (\bibinfo {year} {2023}{\natexlab{b}})}\BibitemShut {NoStop}%
\bibitem [{\citenamefont {Potter}\ \emph {et~al.}(2017)\citenamefont {Potter}, \citenamefont {Byrd}, \citenamefont {Mugler},\ and\ \citenamefont {Sun}}]{Potter2017}%
  \BibitemOpen
  \bibfield  {author} {\bibinfo {author} {\bibfnamefont {G.~D.}\ \bibnamefont {Potter}}, \bibinfo {author} {\bibfnamefont {T.~A.}\ \bibnamefont {Byrd}}, \bibinfo {author} {\bibfnamefont {A.}~\bibnamefont {Mugler}},\ and\ \bibinfo {author} {\bibfnamefont {B.}~\bibnamefont {Sun}},\ }\href {https://doi.org/10.1016/j.bpj.2016.12.045} {\bibfield  {journal} {\bibinfo  {journal} {Biophys. J.}\ }\textbf {\bibinfo {volume} {112}},\ \bibinfo {pages} {795} (\bibinfo {year} {2017})},\ \Eprint {https://arxiv.org/abs/1607.01841} {arXiv:1607.01841} \BibitemShut {NoStop}%
\bibitem [{\citenamefont {Grabowski}\ \emph {et~al.}(2019)\citenamefont {Grabowski}, \citenamefont {Czy{\.{z}}}, \citenamefont {Kocha{\'{n}}czyk},\ and\ \citenamefont {Lipniacki}}]{Grabowski2019}%
  \BibitemOpen
  \bibfield  {author} {\bibinfo {author} {\bibfnamefont {F.}~\bibnamefont {Grabowski}}, \bibinfo {author} {\bibfnamefont {P.}~\bibnamefont {Czy{\.{z}}}}, \bibinfo {author} {\bibfnamefont {M.}~\bibnamefont {Kocha{\'{n}}czyk}},\ and\ \bibinfo {author} {\bibfnamefont {T.}~\bibnamefont {Lipniacki}},\ }\bibfield  {journal} {\bibinfo  {journal} {J. R. Soc. Interface}\ }\textbf {\bibinfo {volume} {16}},\ \href {https://doi.org/10.1098/rsif.2018.0792} {10.1098/rsif.2018.0792} (\bibinfo {year} {2019})\BibitemShut {NoStop}%
\bibitem [{\citenamefont {Cheong}\ \emph {et~al.}(2011)\citenamefont {Cheong}, \citenamefont {Rhee}, \citenamefont {Wang}, \citenamefont {Nemenman},\ and\ \citenamefont {Levchenko}}]{Cheong2011}%
  \BibitemOpen
  \bibfield  {author} {\bibinfo {author} {\bibfnamefont {R.}~\bibnamefont {Cheong}}, \bibinfo {author} {\bibfnamefont {A.}~\bibnamefont {Rhee}}, \bibinfo {author} {\bibfnamefont {C.~J.}\ \bibnamefont {Wang}}, \bibinfo {author} {\bibfnamefont {I.}~\bibnamefont {Nemenman}},\ and\ \bibinfo {author} {\bibfnamefont {A.}~\bibnamefont {Levchenko}},\ }\href {https://doi.org/10.1126/science.1204553} {\bibfield  {journal} {\bibinfo  {journal} {Science (80-. ).}\ }\textbf {\bibinfo {volume} {334}},\ \bibinfo {pages} {354} (\bibinfo {year} {2011})}\BibitemShut {NoStop}%
\bibitem [{\citenamefont {Vazquez-Jimenez}\ and\ \citenamefont {Rodriguez-Gonzalez}(2019)}]{Vazquez-Jimenez2019}%
  \BibitemOpen
  \bibfield  {author} {\bibinfo {author} {\bibfnamefont {A.}~\bibnamefont {Vazquez-Jimenez}}\ and\ \bibinfo {author} {\bibfnamefont {J.}~\bibnamefont {Rodriguez-Gonzalez}},\ }\href {https://doi.org/10.1038/s41598-019-50631-0} {\bibfield  {journal} {\bibinfo  {journal} {Sci. Rep.}\ }\textbf {\bibinfo {volume} {9}},\ \bibinfo {pages} {1} (\bibinfo {year} {2019})}\BibitemShut {NoStop}%
\bibitem [{\citenamefont {Bialek}(2012)}]{Bialek2012}%
  \BibitemOpen
  \bibfield  {author} {\bibinfo {author} {\bibfnamefont {W.}~\bibnamefont {Bialek}},\ }\href@noop {} {\emph {\bibinfo {title} {Biophysics: Searching for Principles}}}\ (\bibinfo  {publisher} {Princeton University Press},\ \bibinfo {year} {2012})\BibitemShut {NoStop}%
\bibitem [{\citenamefont {Bauer}\ \emph {et~al.}(2021)\citenamefont {Bauer}, \citenamefont {Petkova}, \citenamefont {Gregor}, \citenamefont {Wieschaus},\ and\ \citenamefont {Bialek}}]{Bauer2021}%
  \BibitemOpen
  \bibfield  {author} {\bibinfo {author} {\bibfnamefont {M.}~\bibnamefont {Bauer}}, \bibinfo {author} {\bibfnamefont {M.~D.}\ \bibnamefont {Petkova}}, \bibinfo {author} {\bibfnamefont {T.}~\bibnamefont {Gregor}}, \bibinfo {author} {\bibfnamefont {E.~F.}\ \bibnamefont {Wieschaus}},\ and\ \bibinfo {author} {\bibfnamefont {W.}~\bibnamefont {Bialek}},\ }\bibfield  {journal} {\bibinfo  {journal} {Proc. Natl. Acad. Sci. U. S. A.}\ }\textbf {\bibinfo {volume} {118}},\ \href {https://doi.org/10.1073/pnas.2109011118} {10.1073/pnas.2109011118} (\bibinfo {year} {2021})\BibitemShut {NoStop}%
\bibitem [{\citenamefont {Selimkhanov}\ \emph {et~al.}(2014)\citenamefont {Selimkhanov}, \citenamefont {Taylor}, \citenamefont {Yao}, \citenamefont {Pilko}, \citenamefont {Albeck}, \citenamefont {Hoffmann}, \citenamefont {Tsimring},\ and\ \citenamefont {Wollman}}]{Selimkhanov2014}%
  \BibitemOpen
  \bibfield  {author} {\bibinfo {author} {\bibfnamefont {J.}~\bibnamefont {Selimkhanov}}, \bibinfo {author} {\bibfnamefont {B.}~\bibnamefont {Taylor}}, \bibinfo {author} {\bibfnamefont {J.}~\bibnamefont {Yao}}, \bibinfo {author} {\bibfnamefont {A.}~\bibnamefont {Pilko}}, \bibinfo {author} {\bibfnamefont {J.}~\bibnamefont {Albeck}}, \bibinfo {author} {\bibfnamefont {A.}~\bibnamefont {Hoffmann}}, \bibinfo {author} {\bibfnamefont {L.}~\bibnamefont {Tsimring}},\ and\ \bibinfo {author} {\bibfnamefont {R.}~\bibnamefont {Wollman}},\ }\href {https://doi.org/10.1126/science.1254933} {\bibfield  {journal} {\bibinfo  {journal} {Science (80-.).}\ }\textbf {\bibinfo {volume} {346}},\ \bibinfo {pages} {1370} (\bibinfo {year} {2014})}\BibitemShut {NoStop}%
\bibitem [{\citenamefont {Kramer}\ \emph {et~al.}(2022)\citenamefont {Kramer}, \citenamefont {del Castillo},\ and\ \citenamefont {Pelkmans}}]{Kramer2022}%
  \BibitemOpen
  \bibfield  {author} {\bibinfo {author} {\bibfnamefont {B.~A.}\ \bibnamefont {Kramer}}, \bibinfo {author} {\bibfnamefont {J.~S.}\ \bibnamefont {del Castillo}},\ and\ \bibinfo {author} {\bibfnamefont {L.}~\bibnamefont {Pelkmans}},\ }\href {https://doi.org/10.1126/science.abf4062} {\bibfield  {journal} {\bibinfo  {journal} {Science (80-. ).}\ }\textbf {\bibinfo {volume} {377}},\ \bibinfo {pages} {642} (\bibinfo {year} {2022})}\BibitemShut {NoStop}%
\bibitem [{\citenamefont {Suderman}\ and\ \citenamefont {Deeds}(2018)}]{Suderman2018}%
  \BibitemOpen
  \bibfield  {author} {\bibinfo {author} {\bibfnamefont {R.}~\bibnamefont {Suderman}}\ and\ \bibinfo {author} {\bibfnamefont {E.~J.}\ \bibnamefont {Deeds}},\ }\bibfield  {journal} {\bibinfo  {journal} {Interface Focus}\ }\textbf {\bibinfo {volume} {8}},\ \href {https://doi.org/10.1098/rsfs.2018.0039} {10.1098/rsfs.2018.0039} (\bibinfo {year} {2018})\BibitemShut {NoStop}%
\bibitem [{\citenamefont {Elowitz}\ \emph {et~al.}(2002)\citenamefont {Elowitz}, \citenamefont {Levine}, \citenamefont {Siggia},\ and\ \citenamefont {Swain}}]{Elowitz2002}%
  \BibitemOpen
  \bibfield  {author} {\bibinfo {author} {\bibfnamefont {M.~B.}\ \bibnamefont {Elowitz}}, \bibinfo {author} {\bibfnamefont {A.~J.}\ \bibnamefont {Levine}}, \bibinfo {author} {\bibfnamefont {E.~D.}\ \bibnamefont {Siggia}},\ and\ \bibinfo {author} {\bibfnamefont {P.~S.}\ \bibnamefont {Swain}},\ }\href {https://doi.org/10.1126/science.1070919} {\bibfield  {journal} {\bibinfo  {journal} {Science}\ }\textbf {\bibinfo {volume} {297}},\ \bibinfo {pages} {1183} (\bibinfo {year} {2002})}\BibitemShut {NoStop}%
\bibitem [{\citenamefont {Levchenko}\ and\ \citenamefont {Nemenman}(2014)}]{Levchenko2014}%
  \BibitemOpen
  \bibfield  {author} {\bibinfo {author} {\bibfnamefont {A.}~\bibnamefont {Levchenko}}\ and\ \bibinfo {author} {\bibfnamefont {I.}~\bibnamefont {Nemenman}},\ }\href {https://doi.org/10.1016/j.copbio.2014.05.002} {\bibfield  {journal} {\bibinfo  {journal} {Curr. Opin. Biotechnol.}\ }\textbf {\bibinfo {volume} {28}},\ \bibinfo {pages} {156} (\bibinfo {year} {2014})}\BibitemShut {NoStop}%
\bibitem [{\citenamefont {Mitchell}\ and\ \citenamefont {Hoffmann}(2018)}]{Mitchell2018}%
  \BibitemOpen
  \bibfield  {author} {\bibinfo {author} {\bibfnamefont {S.}~\bibnamefont {Mitchell}}\ and\ \bibinfo {author} {\bibfnamefont {A.}~\bibnamefont {Hoffmann}},\ }\href {https://doi.org/10.1016/j.coisb.2017.11.013} {\bibfield  {journal} {\bibinfo  {journal} {Curr. Opin. Syst. Biol.}\ }\textbf {\bibinfo {volume} {8}},\ \bibinfo {pages} {39} (\bibinfo {year} {2018})}\BibitemShut {NoStop}%
\bibitem [{\citenamefont {Tay}\ \emph {et~al.}(2010)\citenamefont {Tay}, \citenamefont {Hughey}, \citenamefont {Lee}, \citenamefont {Lipniacki}, \citenamefont {Quake},\ and\ \citenamefont {Covert}}]{Tay2010}%
  \BibitemOpen
  \bibfield  {author} {\bibinfo {author} {\bibfnamefont {S.}~\bibnamefont {Tay}}, \bibinfo {author} {\bibfnamefont {J.~J.}\ \bibnamefont {Hughey}}, \bibinfo {author} {\bibfnamefont {T.~K.}\ \bibnamefont {Lee}}, \bibinfo {author} {\bibfnamefont {T.}~\bibnamefont {Lipniacki}}, \bibinfo {author} {\bibfnamefont {S.~R.}\ \bibnamefont {Quake}},\ and\ \bibinfo {author} {\bibfnamefont {M.~W.}\ \bibnamefont {Covert}},\ }\href {https://doi.org/10.1038/nature09145} {\bibfield  {journal} {\bibinfo  {journal} {Nature}\ }\textbf {\bibinfo {volume} {466}},\ \bibinfo {pages} {267} (\bibinfo {year} {2010})}\BibitemShut {NoStop}%
\bibitem [{\citenamefont {Vis}\ \emph {et~al.}(2020)\citenamefont {Vis}, \citenamefont {Ito},\ and\ \citenamefont {Hofmann}}]{Vis2020}%
  \BibitemOpen
  \bibfield  {author} {\bibinfo {author} {\bibfnamefont {M.~A.}\ \bibnamefont {Vis}}, \bibinfo {author} {\bibfnamefont {K.}~\bibnamefont {Ito}},\ and\ \bibinfo {author} {\bibfnamefont {S.}~\bibnamefont {Hofmann}},\ }\href {https://doi.org/10.3389/fbioe.2020.00911} {\bibfield  {journal} {\bibinfo  {journal} {Front. Bioeng. Biotechnol.}\ }\textbf {\bibinfo {volume} {8}},\ \bibinfo {pages} {1} (\bibinfo {year} {2020})}\BibitemShut {NoStop}%
\bibitem [{\citenamefont {Goetz}\ \emph {et~al.}(2024)\citenamefont {Goetz}, \citenamefont {Akl},\ and\ \citenamefont {Dixit}}]{Goetz2024}%
  \BibitemOpen
  \bibfield  {author} {\bibinfo {author} {\bibfnamefont {A.}~\bibnamefont {Goetz}}, \bibinfo {author} {\bibfnamefont {H.}~\bibnamefont {Akl}},\ and\ \bibinfo {author} {\bibfnamefont {P.}~\bibnamefont {Dixit}},\ }\href {https://doi.org/10.7554/eLife.87747} {\bibfield  {journal} {\bibinfo  {journal} {Elife}\ }\textbf {\bibinfo {volume} {13}},\ \bibinfo {pages} {1} (\bibinfo {year} {2024})}\BibitemShut {NoStop}%
\bibitem [{\citenamefont {Toettcher}\ \emph {et~al.}(2014)\citenamefont {Toettcher}, \citenamefont {Weiner},\ and\ \citenamefont {Lim}}]{Toettcher2014}%
  \BibitemOpen
  \bibfield  {author} {\bibinfo {author} {\bibfnamefont {J.~E.}\ \bibnamefont {Toettcher}}, \bibinfo {author} {\bibfnamefont {O.~D.}\ \bibnamefont {Weiner}},\ and\ \bibinfo {author} {\bibfnamefont {W.~A.}\ \bibnamefont {Lim}},\ }\href {https://doi.org/10.1016/j.cell.2013.11.004.Using} {\bibfield  {journal} {\bibinfo  {journal} {Nih}\ }\textbf {\bibinfo {volume} {155}},\ \bibinfo {pages} {1422} (\bibinfo {year} {2014})}\BibitemShut {NoStop}%
\bibitem [{\citenamefont {Jetka}\ \emph {et~al.}(2019)\citenamefont {Jetka}, \citenamefont {Niena{\l}towski}, \citenamefont {Winarski}, \citenamefont {B{\l}o{\'{n}}ski},\ and\ \citenamefont {Komorowski}}]{Jetka2019}%
  \BibitemOpen
  \bibfield  {author} {\bibinfo {author} {\bibfnamefont {T.}~\bibnamefont {Jetka}}, \bibinfo {author} {\bibfnamefont {K.}~\bibnamefont {Niena{\l}towski}}, \bibinfo {author} {\bibfnamefont {T.}~\bibnamefont {Winarski}}, \bibinfo {author} {\bibfnamefont {S.}~\bibnamefont {B{\l}o{\'{n}}ski}},\ and\ \bibinfo {author} {\bibfnamefont {M.}~\bibnamefont {Komorowski}},\ }\href {https://doi.org/10.1371/journal.pcbi.1007132} {\bibfield  {journal} {\bibinfo  {journal} {PLoS Comput. Biol.}\ }\textbf {\bibinfo {volume} {15}},\ \bibinfo {pages} {1} (\bibinfo {year} {2019})}\BibitemShut {NoStop}%
\bibitem [{\citenamefont {Kim}\ \emph {et~al.}(2014)\citenamefont {Kim}, \citenamefont {Kim}, \citenamefont {Lee}, \citenamefont {Kim}, \citenamefont {Chang},\ and\ \citenamefont {Heo}}]{Kim2014}%
  \BibitemOpen
  \bibfield  {author} {\bibinfo {author} {\bibfnamefont {N.}~\bibnamefont {Kim}}, \bibinfo {author} {\bibfnamefont {J.~M.}\ \bibnamefont {Kim}}, \bibinfo {author} {\bibfnamefont {M.}~\bibnamefont {Lee}}, \bibinfo {author} {\bibfnamefont {C.~Y.}\ \bibnamefont {Kim}}, \bibinfo {author} {\bibfnamefont {K.~Y.}\ \bibnamefont {Chang}},\ and\ \bibinfo {author} {\bibfnamefont {W.~D.}\ \bibnamefont {Heo}},\ }\href {https://doi.org/10.1016/j.chembiol.2014.05.013} {\bibfield  {journal} {\bibinfo  {journal} {Chem. Biol.}\ }\textbf {\bibinfo {volume} {21}},\ \bibinfo {pages} {903} (\bibinfo {year} {2014})}\BibitemShut {NoStop}%
\bibitem [{\citenamefont {Dessauges}\ \emph {et~al.}(2022)\citenamefont {Dessauges}, \citenamefont {Mikelson}, \citenamefont {Dobrzy\'{n}ski}, \citenamefont {Jacques}, \citenamefont {Frismantiene}, \citenamefont {Gagliardi}, \citenamefont {Khammash},\ and\ \citenamefont {Pertz}}]{Dessauges2022}%
  \BibitemOpen
  \bibfield  {author} {\bibinfo {author} {\bibfnamefont {C.}~\bibnamefont {Dessauges}}, \bibinfo {author} {\bibfnamefont {J.}~\bibnamefont {Mikelson}}, \bibinfo {author} {\bibfnamefont {M.}~\bibnamefont {Dobrzy\'{n}ski}}, \bibinfo {author} {\bibfnamefont {M.}~\bibnamefont {Jacques}}, \bibinfo {author} {\bibfnamefont {A.}~\bibnamefont {Frismantiene}}, \bibinfo {author} {\bibfnamefont {P.~A.}\ \bibnamefont {Gagliardi}}, \bibinfo {author} {\bibfnamefont {M.}~\bibnamefont {Khammash}},\ and\ \bibinfo {author} {\bibfnamefont {O.}~\bibnamefont {Pertz}},\ }\href {https://doi.org/https://doi.org/10.15252/msb.202110670} {\bibfield  {journal} {\bibinfo  {journal} {Molecular Systems Biology}\ }\textbf {\bibinfo {volume} {18}},\ \bibinfo {pages} {e10670} (\bibinfo {year} {2022})}\BibitemShut {NoStop}%
\bibitem [{\citenamefont {de~la Cova}\ \emph {et~al.}(2017)\citenamefont {de~la Cova}, \citenamefont {Townley}, \citenamefont {Regot},\ and\ \citenamefont {Greenwald}}]{Delacova2017}%
  \BibitemOpen
  \bibfield  {author} {\bibinfo {author} {\bibfnamefont {C.}~\bibnamefont {de~la Cova}}, \bibinfo {author} {\bibfnamefont {R.}~\bibnamefont {Townley}}, \bibinfo {author} {\bibfnamefont {S.}~\bibnamefont {Regot}},\ and\ \bibinfo {author} {\bibfnamefont {I.}~\bibnamefont {Greenwald}},\ }\href {https://doi.org/https://doi.org/10.1016/j.devcel.2017.07.014} {\bibfield  {journal} {\bibinfo  {journal} {Dev. Cell}\ }\textbf {\bibinfo {volume} {42}},\ \bibinfo {pages} {542} (\bibinfo {year} {2017})}\BibitemShut {NoStop}%
\bibitem [{\citenamefont {Aoki}\ \emph {et~al.}(2013)\citenamefont {Aoki}, \citenamefont {Kumagai}, \citenamefont {Sakurai}, \citenamefont {Komatsu}, \citenamefont {Fujita}, \citenamefont {Shionyu},\ and\ \citenamefont {Matsuda}}]{Aoki2013}%
  \BibitemOpen
  \bibfield  {author} {\bibinfo {author} {\bibfnamefont {K.}~\bibnamefont {Aoki}}, \bibinfo {author} {\bibfnamefont {Y.}~\bibnamefont {Kumagai}}, \bibinfo {author} {\bibfnamefont {A.}~\bibnamefont {Sakurai}}, \bibinfo {author} {\bibfnamefont {N.}~\bibnamefont {Komatsu}}, \bibinfo {author} {\bibfnamefont {Y.}~\bibnamefont {Fujita}}, \bibinfo {author} {\bibfnamefont {C.}~\bibnamefont {Shionyu}},\ and\ \bibinfo {author} {\bibfnamefont {M.}~\bibnamefont {Matsuda}},\ }\href {https://doi.org/10.1016/j.molcel.2013.09.015} {\bibfield  {journal} {\bibinfo  {journal} {Mol. Cell}\ }\textbf {\bibinfo {volume} {52}},\ \bibinfo {pages} {529} (\bibinfo {year} {2013})}\BibitemShut {NoStop}%
\bibitem [{\citenamefont {Ish-Horowicz}\ and\ \citenamefont {Reid}(2017)}]{Ish-Horowicz2017}%
  \BibitemOpen
  \bibfield  {author} {\bibinfo {author} {\bibfnamefont {J.}~\bibnamefont {Ish-Horowicz}}\ and\ \bibinfo {author} {\bibfnamefont {J.}~\bibnamefont {Reid}},\ }\bibfield  {journal} {\bibinfo  {journal} {bioRxiv}\ }\href {https://doi.org/10.1101/132647} {10.1101/132647} (\bibinfo {year} {2017})\BibitemShut {NoStop}%
\bibitem [{\citenamefont {Hernández}\ and\ \citenamefont {Samengo}(2022)}]{Hernandez2022}%
  \BibitemOpen
  \bibfield  {author} {\bibinfo {author} {\bibfnamefont {D.~G.}\ \bibnamefont {Hernández}}\ and\ \bibinfo {author} {\bibfnamefont {I.}~\bibnamefont {Samengo}},\ }\bibfield  {journal} {\bibinfo  {journal} {Entropy}\ }\textbf {\bibinfo {volume} {24}},\ \href {https://doi.org/10.3390/e24010125} {10.3390/e24010125} (\bibinfo {year} {2022})\BibitemShut {NoStop}%
\bibitem [{\citenamefont {Wilson}\ \emph {et~al.}(2017)\citenamefont {Wilson}, \citenamefont {Ravindran}, \citenamefont {Lim},\ and\ \citenamefont {Toettcher}}]{Wilson2017}%
  \BibitemOpen
  \bibfield  {author} {\bibinfo {author} {\bibfnamefont {M.~Z.}\ \bibnamefont {Wilson}}, \bibinfo {author} {\bibfnamefont {P.~T.}\ \bibnamefont {Ravindran}}, \bibinfo {author} {\bibfnamefont {W.~A.}\ \bibnamefont {Lim}},\ and\ \bibinfo {author} {\bibfnamefont {J.~E.}\ \bibnamefont {Toettcher}},\ }\href {https://doi.org/10.1016/j.molcel.2017.07.016} {\bibfield  {journal} {\bibinfo  {journal} {Mol. Cell}\ }\textbf {\bibinfo {volume} {67}},\ \bibinfo {pages} {757} (\bibinfo {year} {2017})}\BibitemShut {NoStop}%
\bibitem [{\citenamefont {Tka{\v{c}}ik}\ \emph {et~al.}(2008)\citenamefont {Tka{\v{c}}ik}, \citenamefont {Callan},\ and\ \citenamefont {Bialek}}]{Tkacik2008}%
  \BibitemOpen
  \bibfield  {author} {\bibinfo {author} {\bibfnamefont {G.}~\bibnamefont {Tka{\v{c}}ik}}, \bibinfo {author} {\bibfnamefont {C.~G.}\ \bibnamefont {Callan}},\ and\ \bibinfo {author} {\bibfnamefont {W.}~\bibnamefont {Bialek}},\ }\href {https://doi.org/10.1073/pnas.0806077105} {\bibfield  {journal} {\bibinfo  {journal} {Proc. Natl. Acad. Sci. U. S. A.}\ }\textbf {\bibinfo {volume} {105}},\ \bibinfo {pages} {12265} (\bibinfo {year} {2008})}\BibitemShut {NoStop}%
\bibitem [{\citenamefont {Stanoev}\ \emph {et~al.}(2020)\citenamefont {Stanoev}, \citenamefont {Nandan},\ and\ \citenamefont {Koseska}}]{Stanoev2020}%
  \BibitemOpen
  \bibfield  {author} {\bibinfo {author} {\bibfnamefont {A.}~\bibnamefont {Stanoev}}, \bibinfo {author} {\bibfnamefont {A.~P.}\ \bibnamefont {Nandan}},\ and\ \bibinfo {author} {\bibfnamefont {A.}~\bibnamefont {Koseska}},\ }\href {https://doi.org/10.15252/msb.20198870} {\bibfield  {journal} {\bibinfo  {journal} {Mol. Syst. Biol.}\ }\textbf {\bibinfo {volume} {16}},\ \bibinfo {pages} {1} (\bibinfo {year} {2020})}\BibitemShut {NoStop}%
\bibitem [{\citenamefont {Aikin}\ \emph {et~al.}(2020)\citenamefont {Aikin}, \citenamefont {Peterson}, \citenamefont {Pokrass}, \citenamefont {Clark},\ and\ \citenamefont {Regot}}]{Aikin2020}%
  \BibitemOpen
  \bibfield  {author} {\bibinfo {author} {\bibfnamefont {T.~J.}\ \bibnamefont {Aikin}}, \bibinfo {author} {\bibfnamefont {A.~F.}\ \bibnamefont {Peterson}}, \bibinfo {author} {\bibfnamefont {M.~J.}\ \bibnamefont {Pokrass}}, \bibinfo {author} {\bibfnamefont {H.~R.}\ \bibnamefont {Clark}},\ and\ \bibinfo {author} {\bibfnamefont {S.}~\bibnamefont {Regot}},\ }\href {https://doi.org/10.7554/ELIFE.60541} {\bibfield  {journal} {\bibinfo  {journal} {Elife}\ }\textbf {\bibinfo {volume} {9}},\ \bibinfo {pages} {1} (\bibinfo {year} {2020})}\BibitemShut {NoStop}%
\bibitem [{\citenamefont {Gagliardi}\ \emph {et~al.}(2021)\citenamefont {Gagliardi}, \citenamefont {Dobrzy{\'{n}}ski}, \citenamefont {Jacques}, \citenamefont {Dessauges}, \citenamefont {Ender}, \citenamefont {Blum}, \citenamefont {Hughes}, \citenamefont {Cohen},\ and\ \citenamefont {Pertz}}]{Gagliardi2021}%
  \BibitemOpen
  \bibfield  {author} {\bibinfo {author} {\bibfnamefont {P.~A.}\ \bibnamefont {Gagliardi}}, \bibinfo {author} {\bibfnamefont {M.}~\bibnamefont {Dobrzy{\'{n}}ski}}, \bibinfo {author} {\bibfnamefont {M.~A.}\ \bibnamefont {Jacques}}, \bibinfo {author} {\bibfnamefont {C.}~\bibnamefont {Dessauges}}, \bibinfo {author} {\bibfnamefont {P.}~\bibnamefont {Ender}}, \bibinfo {author} {\bibfnamefont {Y.}~\bibnamefont {Blum}}, \bibinfo {author} {\bibfnamefont {R.~M.}\ \bibnamefont {Hughes}}, \bibinfo {author} {\bibfnamefont {A.~R.}\ \bibnamefont {Cohen}},\ and\ \bibinfo {author} {\bibfnamefont {O.}~\bibnamefont {Pertz}},\ }\href {https://doi.org/10.1016/j.devcel.2021.05.007} {\bibfield  {journal} {\bibinfo  {journal} {Dev. Cell}\ }\textbf {\bibinfo {volume} {56}},\ \bibinfo {pages} {1712} (\bibinfo {year} {2021})}\BibitemShut {NoStop}%
\bibitem [{\citenamefont {Ender}\ \emph {et~al.}(2022)\citenamefont {Ender}, \citenamefont {Gagliardi}, \citenamefont {Dobrzy{\'{n}}ski}, \citenamefont {Frismantiene}, \citenamefont {Dessauges}, \citenamefont {H{\"{o}}hener}, \citenamefont {Jacques}, \citenamefont {Cohen},\ and\ \citenamefont {Pertz}}]{Ender2022}%
  \BibitemOpen
  \bibfield  {author} {\bibinfo {author} {\bibfnamefont {P.}~\bibnamefont {Ender}}, \bibinfo {author} {\bibfnamefont {P.~A.}\ \bibnamefont {Gagliardi}}, \bibinfo {author} {\bibfnamefont {M.}~\bibnamefont {Dobrzy{\'{n}}ski}}, \bibinfo {author} {\bibfnamefont {A.}~\bibnamefont {Frismantiene}}, \bibinfo {author} {\bibfnamefont {C.}~\bibnamefont {Dessauges}}, \bibinfo {author} {\bibfnamefont {T.}~\bibnamefont {H{\"{o}}hener}}, \bibinfo {author} {\bibfnamefont {M.~A.}\ \bibnamefont {Jacques}}, \bibinfo {author} {\bibfnamefont {A.~R.}\ \bibnamefont {Cohen}},\ and\ \bibinfo {author} {\bibfnamefont {O.}~\bibnamefont {Pertz}},\ }\href {https://doi.org/10.1016/j.devcel.2022.08.008} {\bibfield  {journal} {\bibinfo  {journal} {Dev. Cell}\ }\textbf {\bibinfo {volume} {57}},\ \bibinfo {pages} {2153} (\bibinfo {year} {2022})}\BibitemShut {NoStop}%
\bibitem [{\citenamefont {Kraskov}\ \emph {et~al.}(2004)\citenamefont {Kraskov}, \citenamefont {St{\"{o}}gbauer},\ and\ \citenamefont {Grassberger}}]{Kraskov2004}%
  \BibitemOpen
  \bibfield  {author} {\bibinfo {author} {\bibfnamefont {A.}~\bibnamefont {Kraskov}}, \bibinfo {author} {\bibfnamefont {H.}~\bibnamefont {St{\"{o}}gbauer}},\ and\ \bibinfo {author} {\bibfnamefont {P.}~\bibnamefont {Grassberger}},\ }\href {https://doi.org/10.1103/PhysRevE.69.066138} {\bibfield  {journal} {\bibinfo  {journal} {Phys. Rev. E - Stat. Physics, Plasmas, Fluids, Relat. Interdiscip. Top.}\ }\textbf {\bibinfo {volume} {69}},\ \bibinfo {pages} {16} (\bibinfo {year} {2004})}\BibitemShut {NoStop}%
\bibitem [{\citenamefont {Tang}\ \emph {et~al.}(2021)\citenamefont {Tang}, \citenamefont {Adelaja}, \citenamefont {Ye}, \citenamefont {Deeds}, \citenamefont {Wollman},\ and\ \citenamefont {Hoffmann}}]{Tang2021}%
  \BibitemOpen
  \bibfield  {author} {\bibinfo {author} {\bibfnamefont {Y.}~\bibnamefont {Tang}}, \bibinfo {author} {\bibfnamefont {A.}~\bibnamefont {Adelaja}}, \bibinfo {author} {\bibfnamefont {F.~X.}\ \bibnamefont {Ye}}, \bibinfo {author} {\bibfnamefont {E.}~\bibnamefont {Deeds}}, \bibinfo {author} {\bibfnamefont {R.}~\bibnamefont {Wollman}},\ and\ \bibinfo {author} {\bibfnamefont {A.}~\bibnamefont {Hoffmann}},\ }\href {https://doi.org/10.1038/s41467-021-21562-0} {\bibfield  {journal} {\bibinfo  {journal} {Nat. Commun.}\ }\textbf {\bibinfo {volume} {12}},\ \bibinfo {pages} {1} (\bibinfo {year} {2021})}\BibitemShut {NoStop}%
\bibitem [{\citenamefont {Hahn}\ \emph {et~al.}(2023)\citenamefont {Hahn}, \citenamefont {Walczak},\ and\ \citenamefont {Mora}}]{Hahn2023}%
  \BibitemOpen
  \bibfield  {author} {\bibinfo {author} {\bibfnamefont {L.}~\bibnamefont {Hahn}}, \bibinfo {author} {\bibfnamefont {A.~M.}\ \bibnamefont {Walczak}},\ and\ \bibinfo {author} {\bibfnamefont {T.}~\bibnamefont {Mora}},\ }\href {https://doi.org/10.1103/PhysRevLett.131.128401} {\bibfield  {journal} {\bibinfo  {journal} {Phys. Rev. Lett.}\ }\textbf {\bibinfo {volume} {131}},\ \bibinfo {pages} {128401} (\bibinfo {year} {2023})}\BibitemShut {NoStop}%
\bibitem [{\citenamefont {Mikelson}\ and\ \citenamefont {Khammash}(2020)}]{Mikelson2020}%
  \BibitemOpen
  \bibfield  {author} {\bibinfo {author} {\bibfnamefont {J.}~\bibnamefont {Mikelson}}\ and\ \bibinfo {author} {\bibfnamefont {M.}~\bibnamefont {Khammash}},\ }\href {https://doi.org/10.1371/journal.pcbi.1008264} {\bibfield  {journal} {\bibinfo  {journal} {PLoS Comput. Biol.}\ }\textbf {\bibinfo {volume} {16}},\ \bibinfo {pages} {1} (\bibinfo {year} {2020})}\BibitemShut {NoStop}%
\end{thebibliography}%
\end{document}


\title{Supplemental Material to the article: \\
Single cells can resolve graded stimuli}

\maketitle

\begin{figure*}[h!]
\centering
\includegraphics[width=0.5 \linewidth]{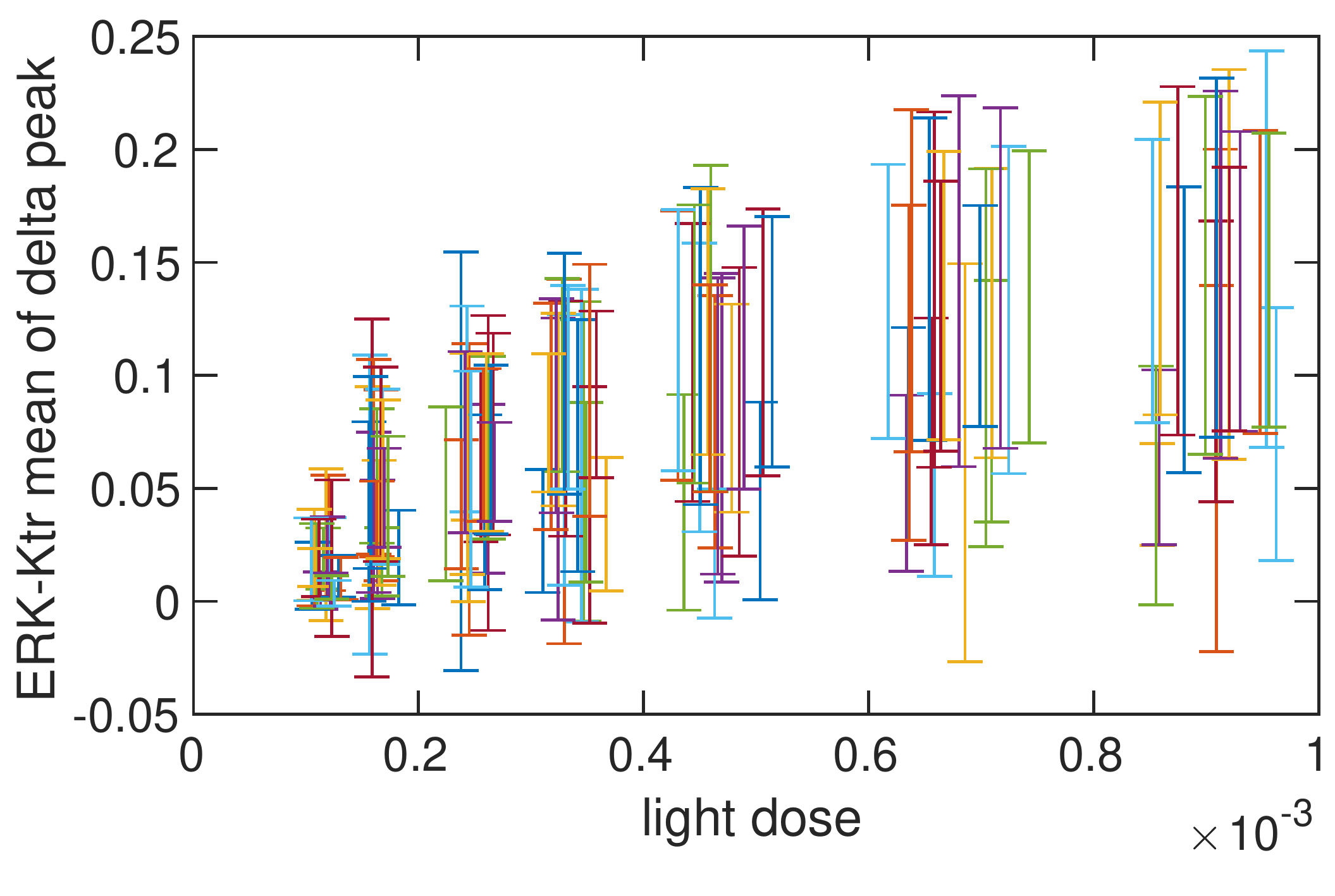}
\caption{\textbf{Dose-responses of all cells} from one stage position in the main dataset (20 repeats). The bars are centered at the mean ERK response value across the 20 repeats to the measured light doses, and the error bars depict the standard deviation.}
\label{fig:SI-DoseRespErrorBarsAllOptoGranPos4}
\end{figure*}

\begin{figure*}[h!]
\centering
\includegraphics[width=0.5 \linewidth]{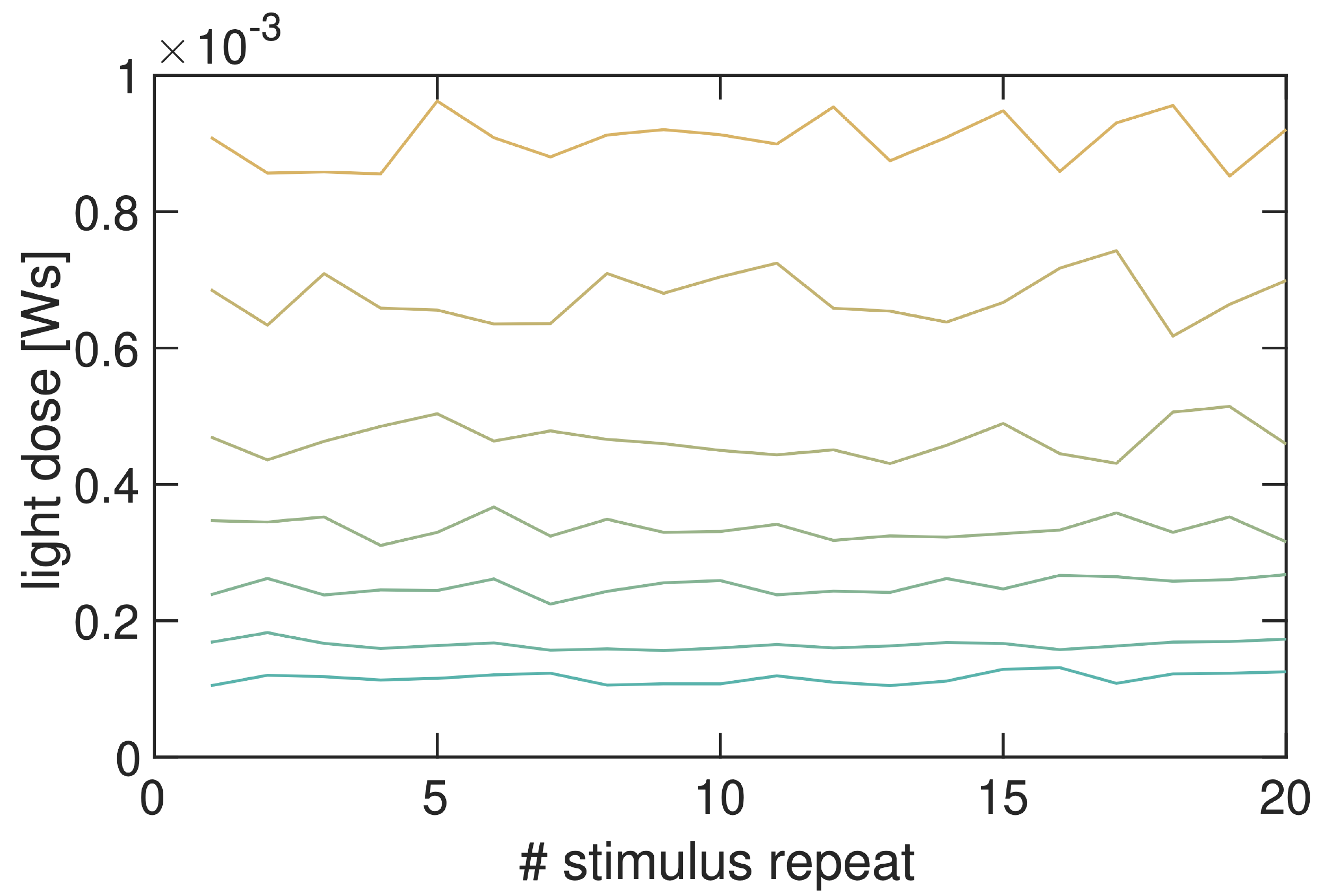}
\caption{\textbf{The variability between repeats of light doses.} Each of the 7 light doses is represented by a line connecting the 20 repeats of the dose. }
\label{fig:SI-InputDosesVariability}
\end{figure*}

\begin{figure*}[h!]
\centering
\includegraphics[width=0.5 \linewidth]{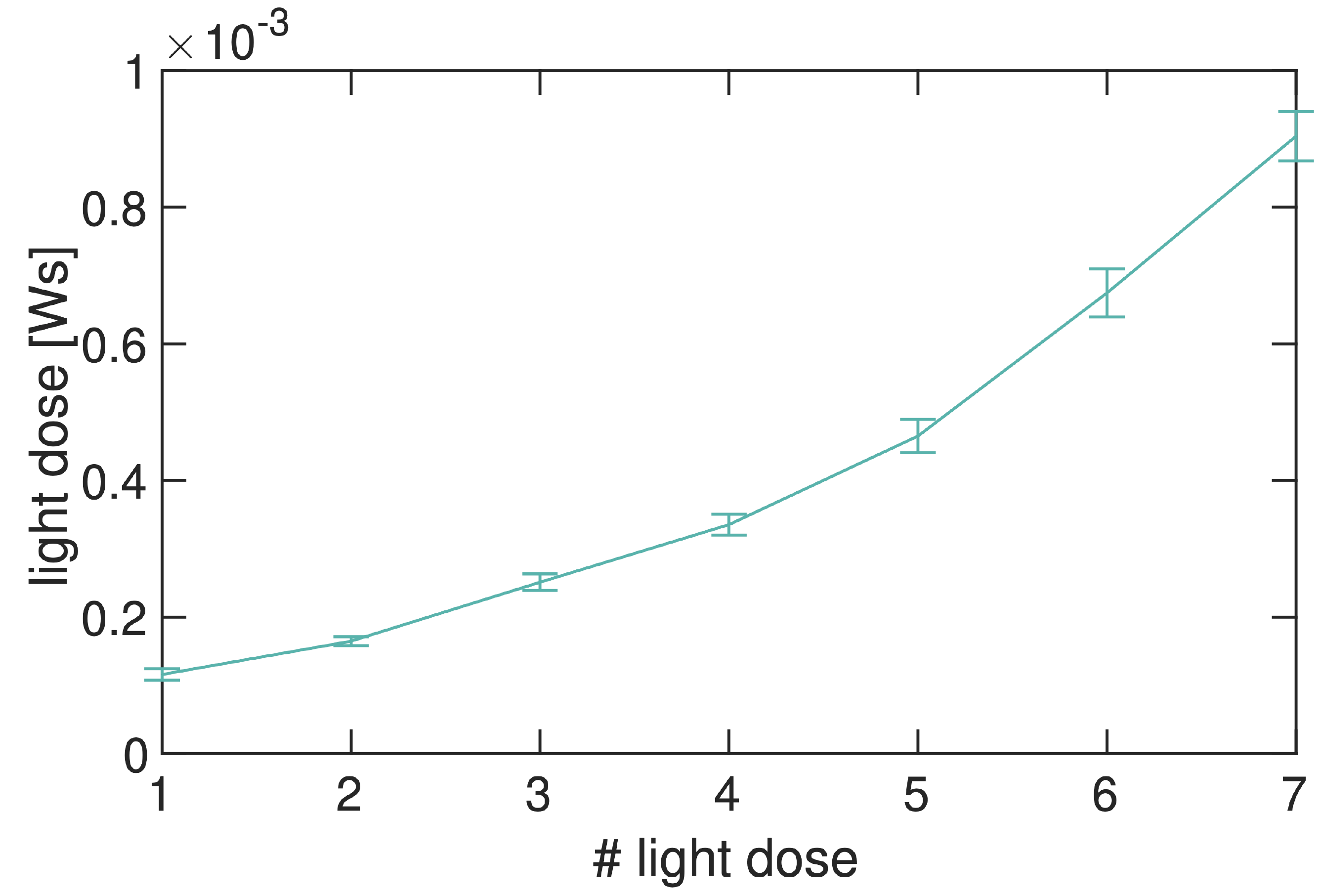}
\caption{\textbf{The average measured light doses}, connected by a line. The error bars show the standard deviation for the 20 repeats of each light dose. }
\label{fig:SI-InputDosesAvg}
\end{figure*}

\begin{figure*}[h!]
\centering
\includegraphics[width=\linewidth]{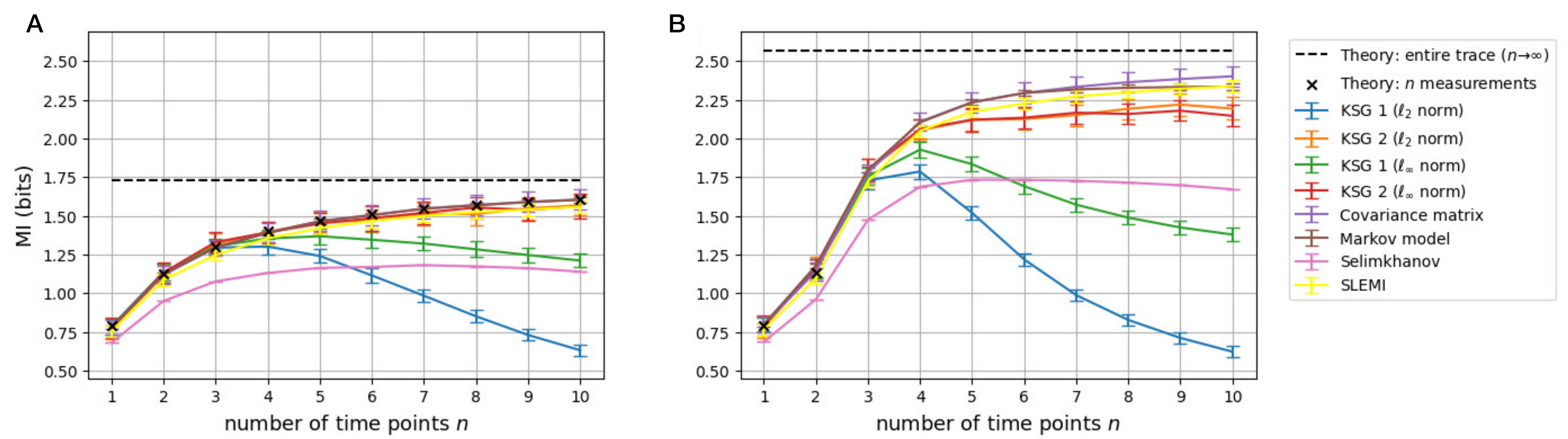}
\caption{\textbf{Benchmarking of mutual information estimators}, plots from~\cite{Hahn2023} for overdamped (a) and underdamped (b) dynamics, here with the addition of SLEMI algorithm. The plots demonstrate that SLEMI over-performs commonly used estimators. Number of samples N= 20000.} 
\label{fig:SI-BenchmarkingMIEstimators}
\end{figure*}

\begin{figure*}[h!]
\centering
\includegraphics[width= \linewidth]{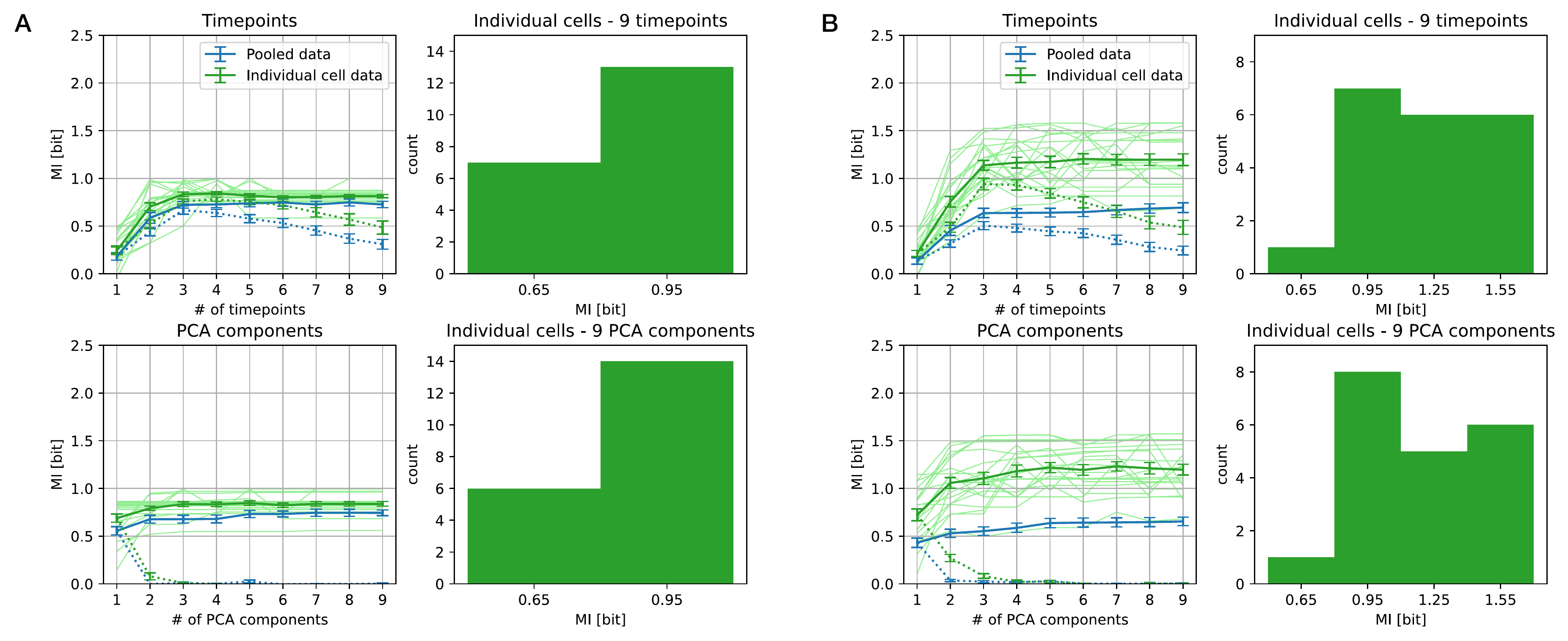}
\caption{\textbf{Mutual information for different number of input states.} Mutual information estimated for the cells of the main dataset with 20 repeats, in panel (A) taking into account only the first (weakest) and the last (7th, strongest) stimulus, while in (B) the fourth (mid-range) stimulus is additionally taken into account. }
\label{fig:SI-NoOfCond2then3}
\end{figure*}

\begin{figure*}[h!]
\centering
\includegraphics[width= 0.8 \linewidth]{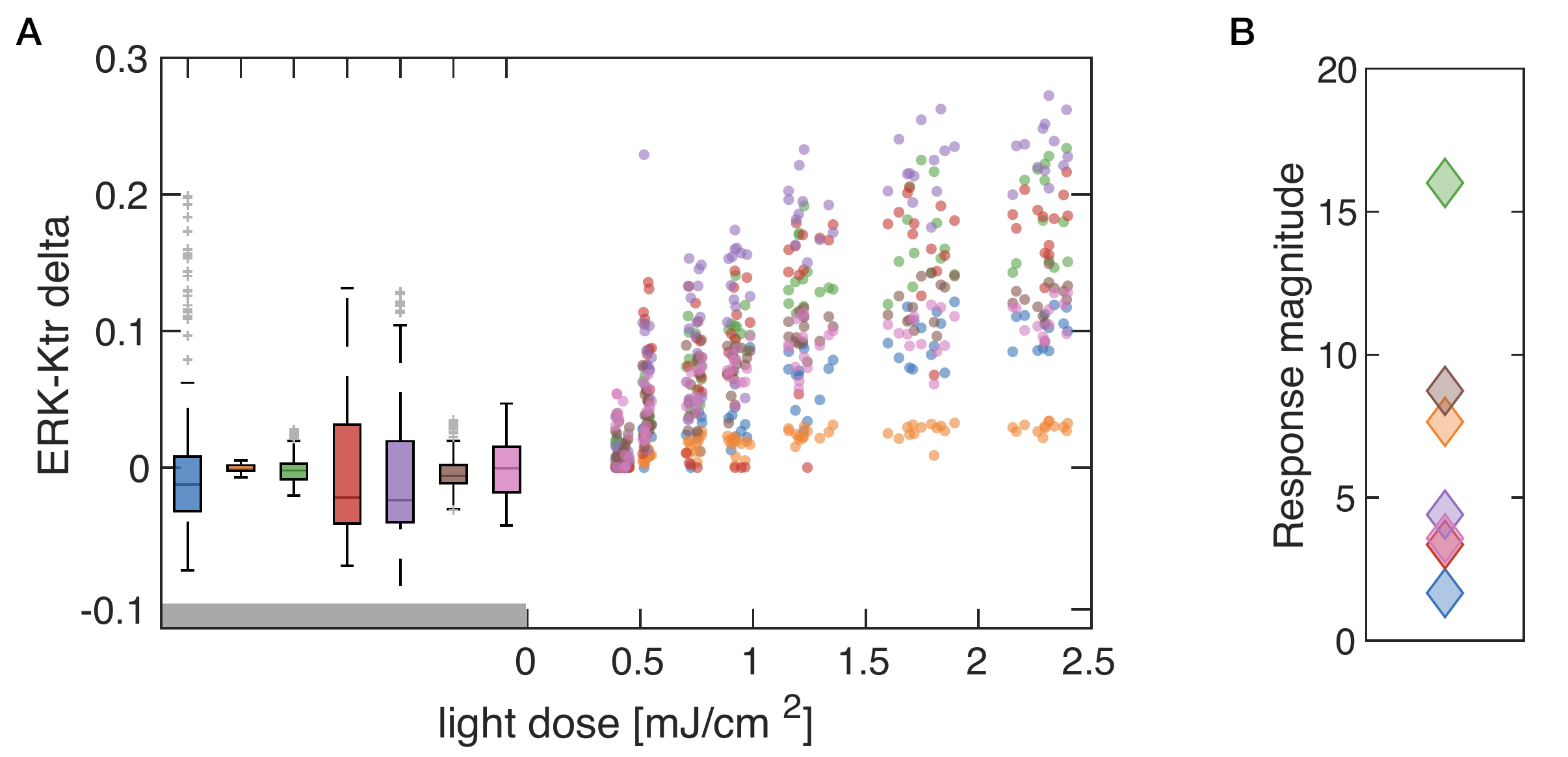}
\caption{\textbf{Comparison between the endogenous ERK fluctuations and ERK response to light stimuli} for all cells from one stage position of the main dataset with 20 repeats (A). In (B), response magnitude is calculated as the ratio between the maximal response to the stimulus and the standard deviation of the endogenous fluctuations.}
\label{fig:SI-AllCellsEndFluctDoseResp}
\end{figure*}

\begin{figure*}[h!]
\centering
\includegraphics[width=\linewidth]{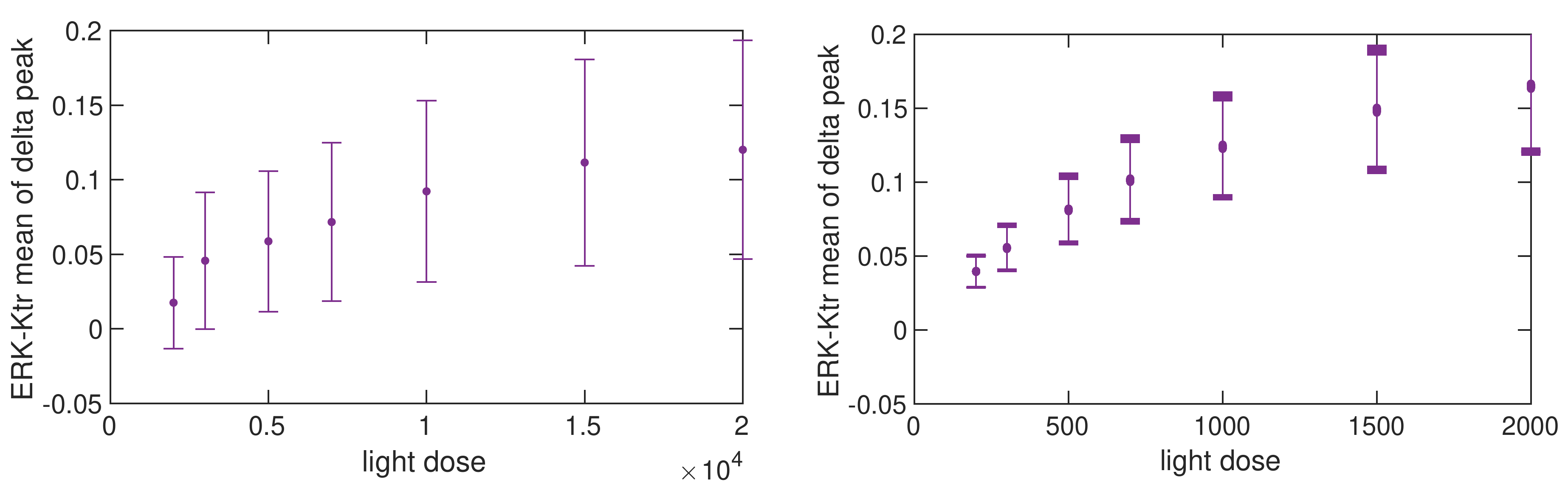}
\caption{\textbf{Experimental (left) and simulated (right) dose-responses.} The dose-responses in the simulations resemble the experimental ones. Note that the relationship between the simulated light doses corresponds to the experimental setting, not their magnitude displayed the plot.}
\label{fig:SI-DoseRespSimExp}
\end{figure*}

\begin{figure*}[h!]
\includegraphics[width= 0.7\linewidth]{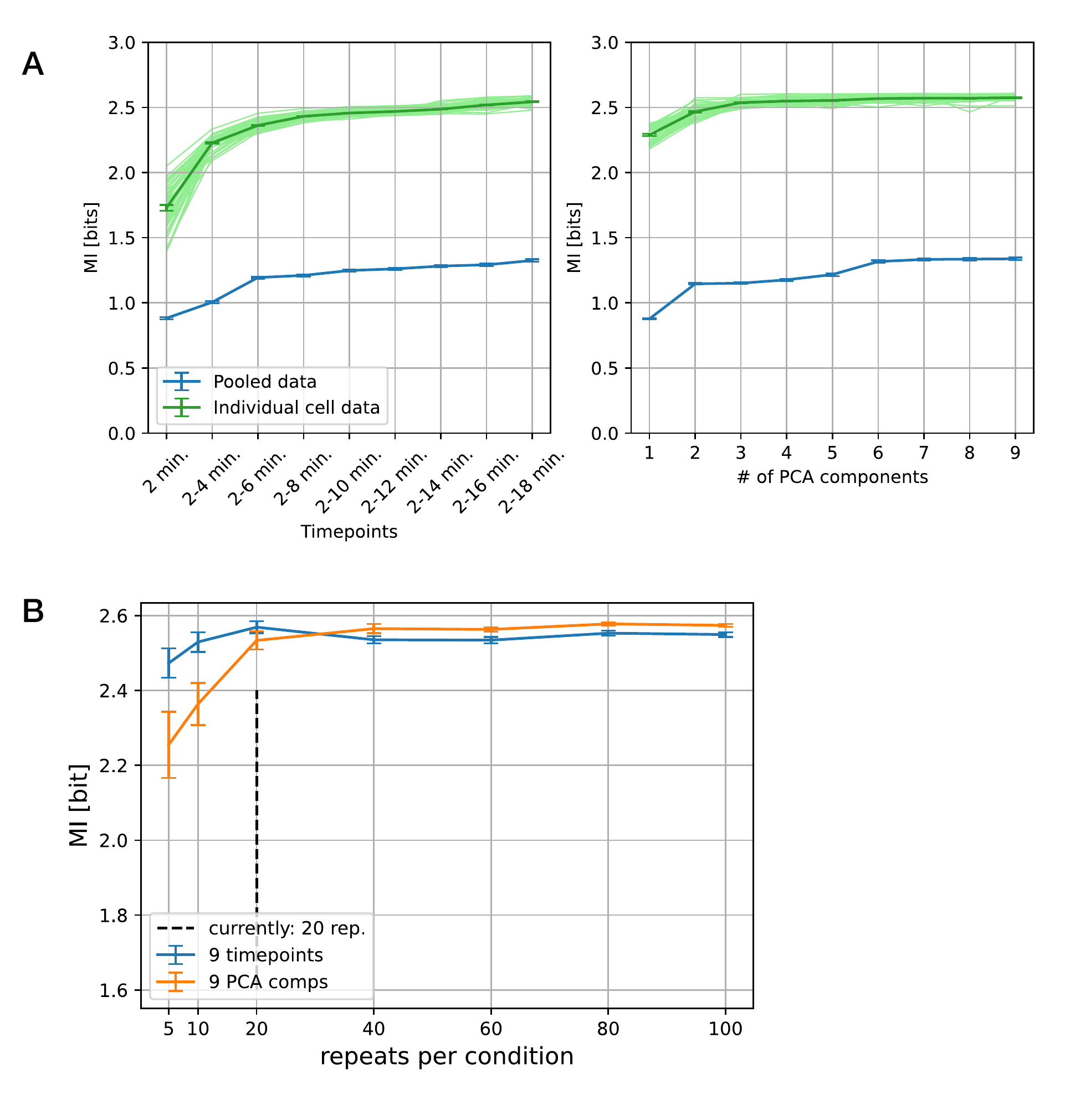}
\caption{\textbf{Simulations of optogenetic activation of the MAPK cascade in conditions of low noise.} Analogously to the Fig.4 in the main text, the trajectories resulting from the simulations described in the main text were modified by adding 0.0001 Gaussian white noise. \textbf{(A)}  Mutual information contained in the simulated cells' ERK response trajectories was calculated using the time point feature reduction (left) and PCA (right). The results from pooled cell responses is shown in blue, and the results from single cell responses in green (individual cells in light green, the average of single cell responses in dark green). \textbf{(B)}~The estimation of mutual information in simulated trajectories for different number of repeats per condition. The information is estimated using the first 9 time points (blue) and PCA (orange). The value of mutual information increases with the increasing sample size, stabilizing at around 20 repeats. The black dashed line is drawn at 20 repeats per condition, the current experimental limit. }
\label{fig:simulations}
\end{figure*}

\begin{figure*}[h!]
\centering
\includegraphics[width= \linewidth]{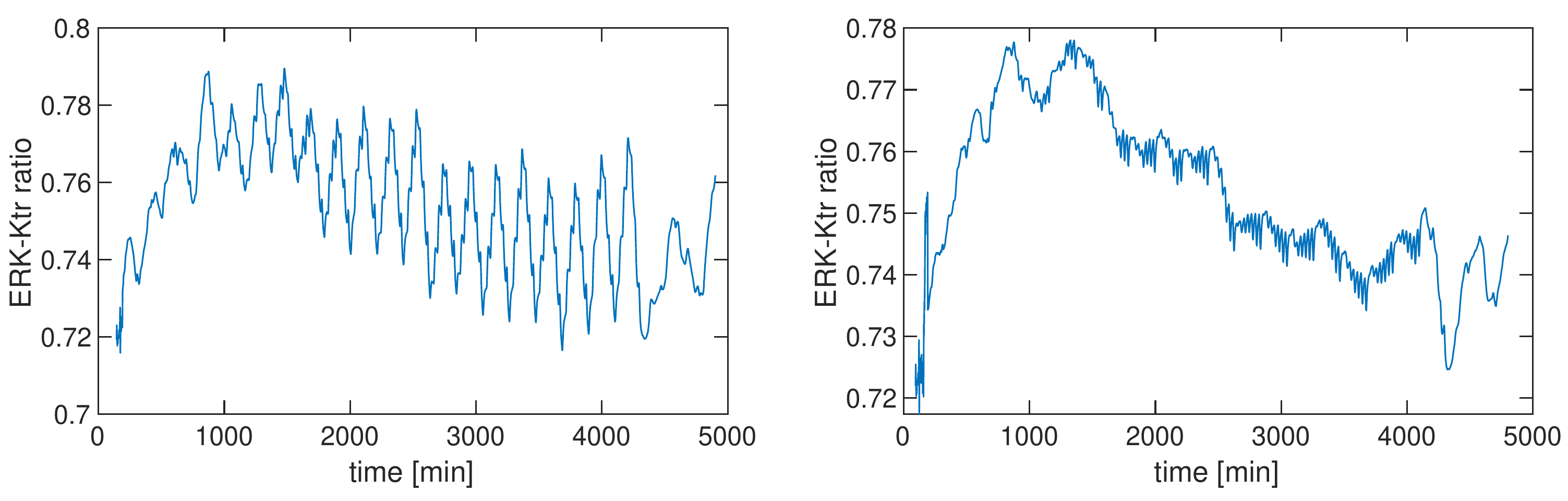}
\caption{\textbf{Trend of a trajectory}, from detrending with a moving averaging window of 50 frames (left) and 100 frames (right).}
\label{fig:SI-TrendMovAvgWindow50and100fr}
\end{figure*}

\begin{figure*}[h!]
\centering
\includegraphics[width=\linewidth]{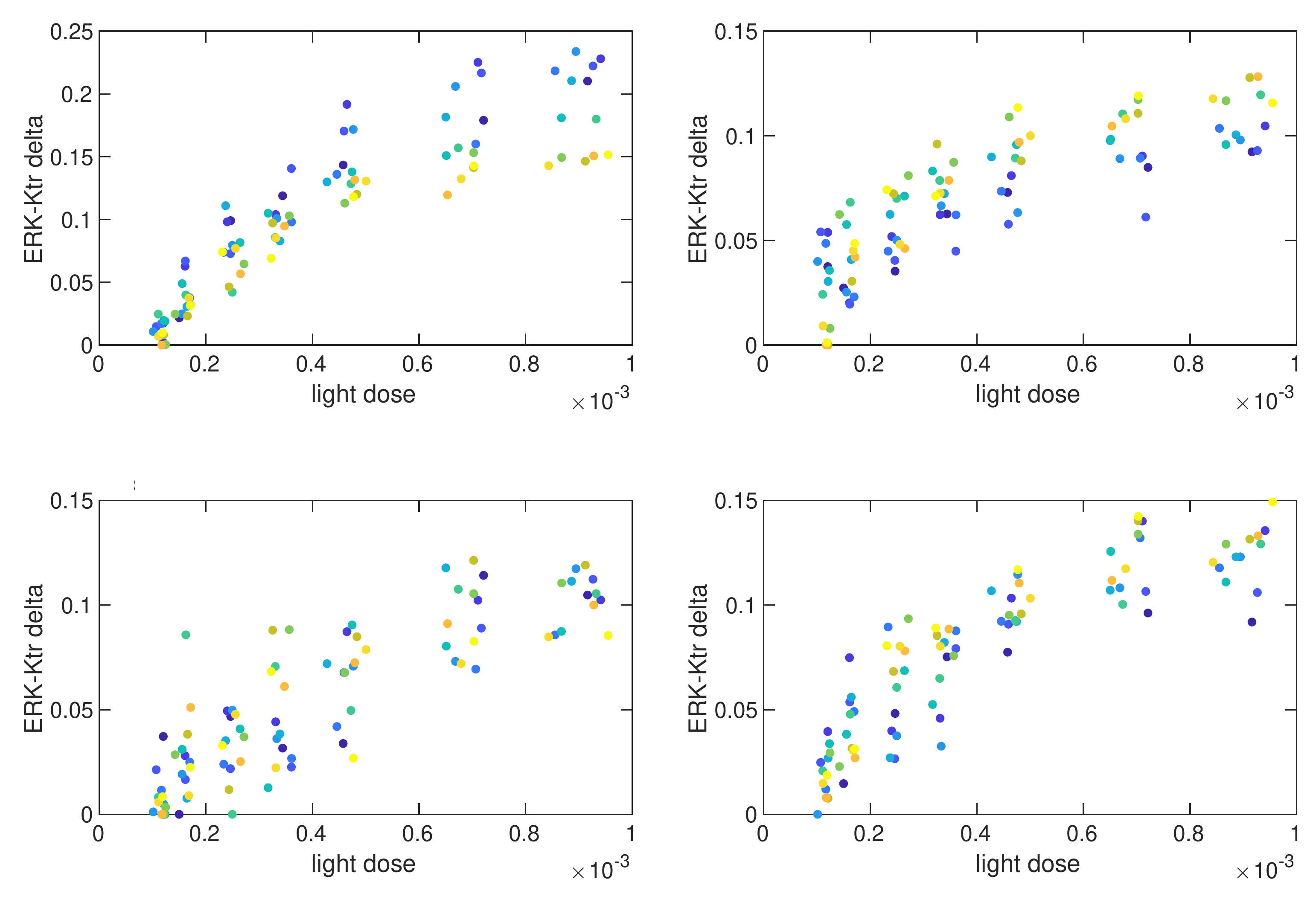}
\caption{\textbf{Some cells could experience a drift in their encoding scheme.} Dose-responses of 4 different cells from the main dataset (20 repeats) and same stage position (i.e. receiving identical light inputs). The marker color codes for increasing repeat number (blue to yellow). Cells in top panels display opposing drifts in reaction to the light stimuli, while the bottom cells do not display a clear drift. The trajectories are detrended with a moving averaging window of 100 frames (3.3h).}
\label{fig:SI-DoseResponsesBluetoYellow}
\end{figure*}

\begin{figure*}[h!]
\centering
\includegraphics[width=\linewidth]{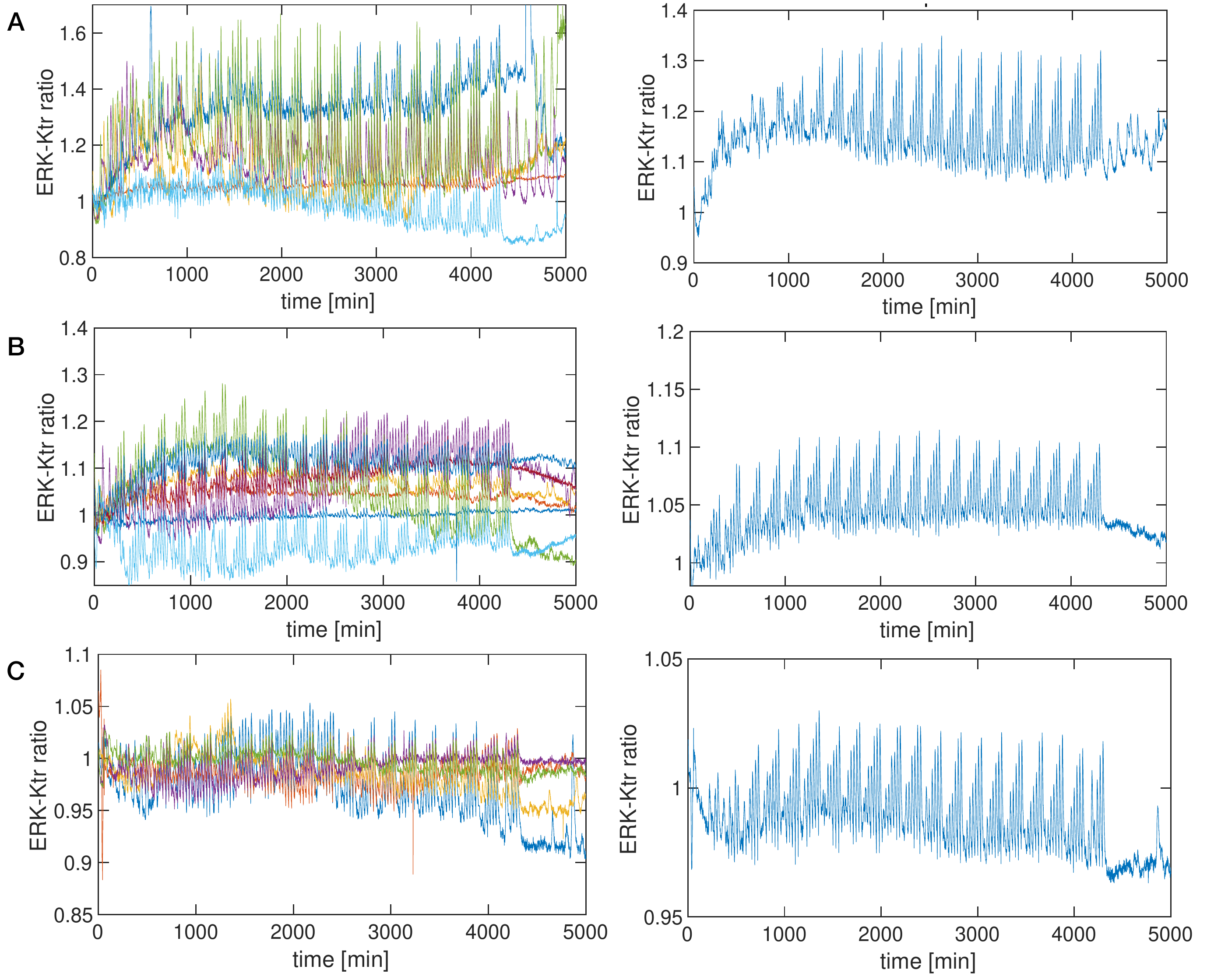}
\caption{\textbf{Raw ERK-KTR ratio trajectories} (left) and their population average (right) of the main dataset with 20 repeats. Panels (A) to (C) depict cells in three different stage positions that were selected for further analysis.}
\label{fig:SI-SI-RawTrajsAndPopAvg}
\end{figure*}

\begin{figure*}[h!]
\centering
\includegraphics[width=\linewidth]{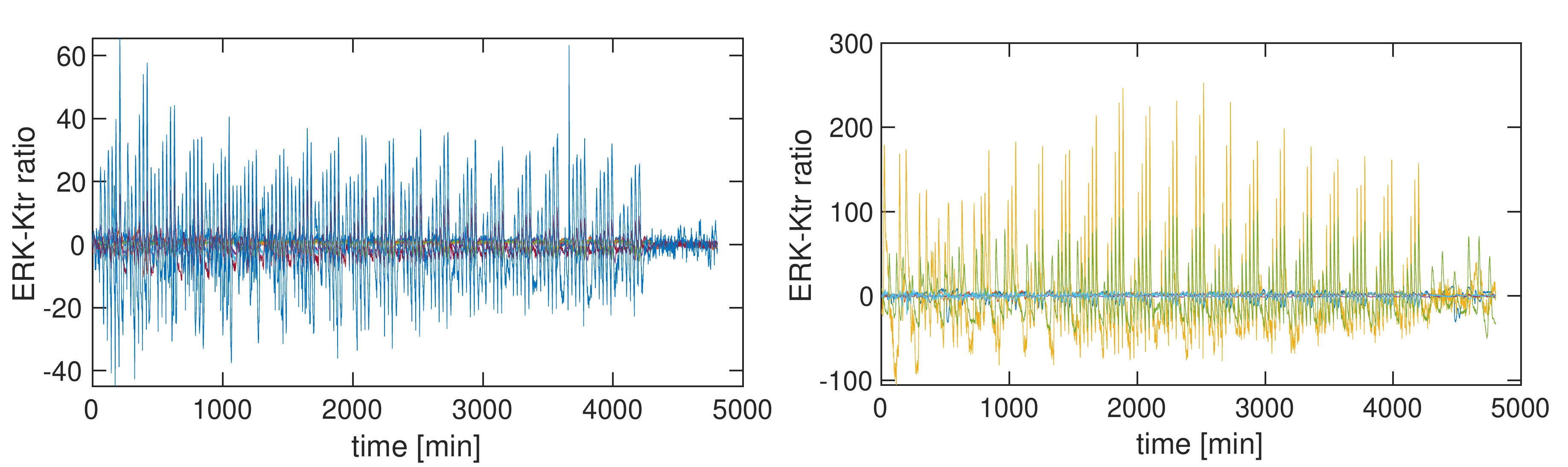}
\caption{\textbf{Detrended trajectories} from two stage positions of the main dataset with 20 repeats, here detrended with a moving averaging window of 100 frames (3.3h).}
\label{fig:SI-DetrendedOptoGran1and4}
\end{figure*}

\begin{figure*}[h!]
\centering
\includegraphics[width=\linewidth]{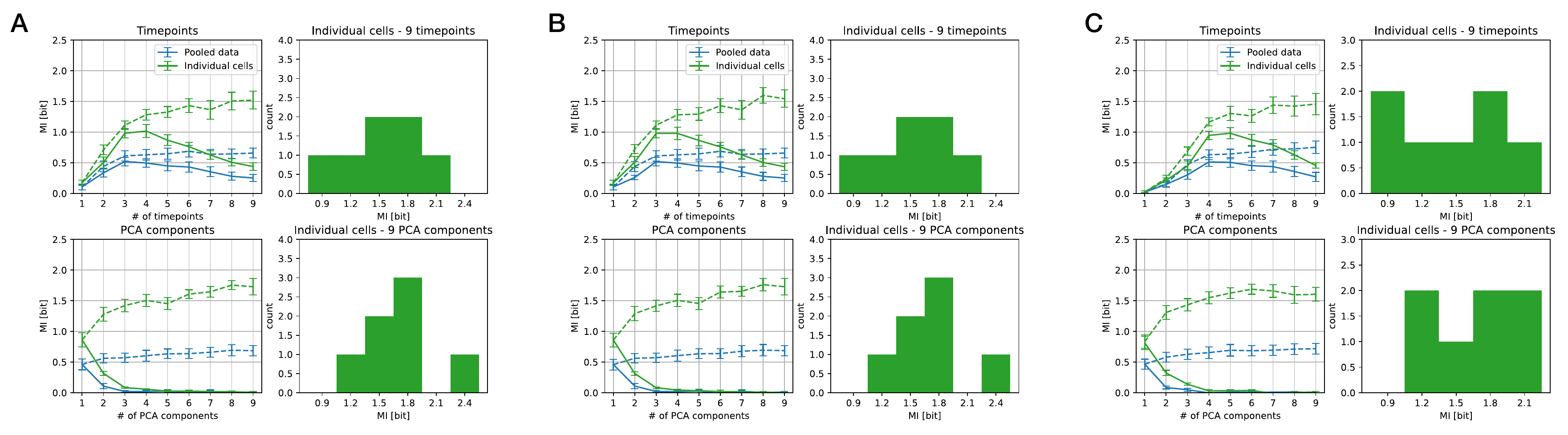}
\caption{\textbf{The effect of detrending.} Mutual information estimated using raw trajectories as input (A), trajectories detrended using a 4th order polynomial (B) and a moving averaging window of 100 frames (3.3h) (C). Time-point based feature reduction is shown at the top, and PCA at the bottom of each panel. The mutual information estimate shows negligible global differences between the three approaches.}
\label{fig:SI-EffectOfDetrending}
\end{figure*}

\begin{figure*}[h!]
\centering
\includegraphics[width=0.6 \linewidth]{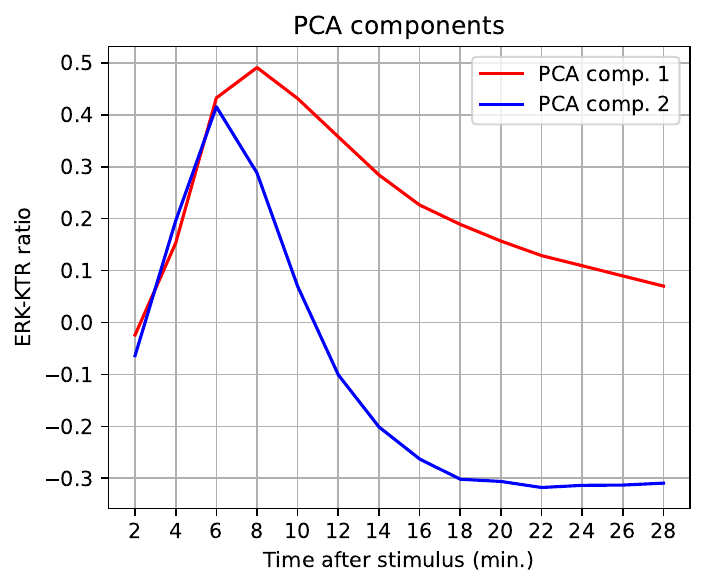}
\caption{\textbf{Principal component analysis.} First two principal components of the ensemble of the response trajectories from the main dataset with 20 stimulation repeats per condition.}
\label{fig:SI-PCAFirstTwoModes}
\end{figure*}

\FloatBarrier

\begin{video}[h!]
    \centering
    \includegraphics[width=0.5 \linewidth]{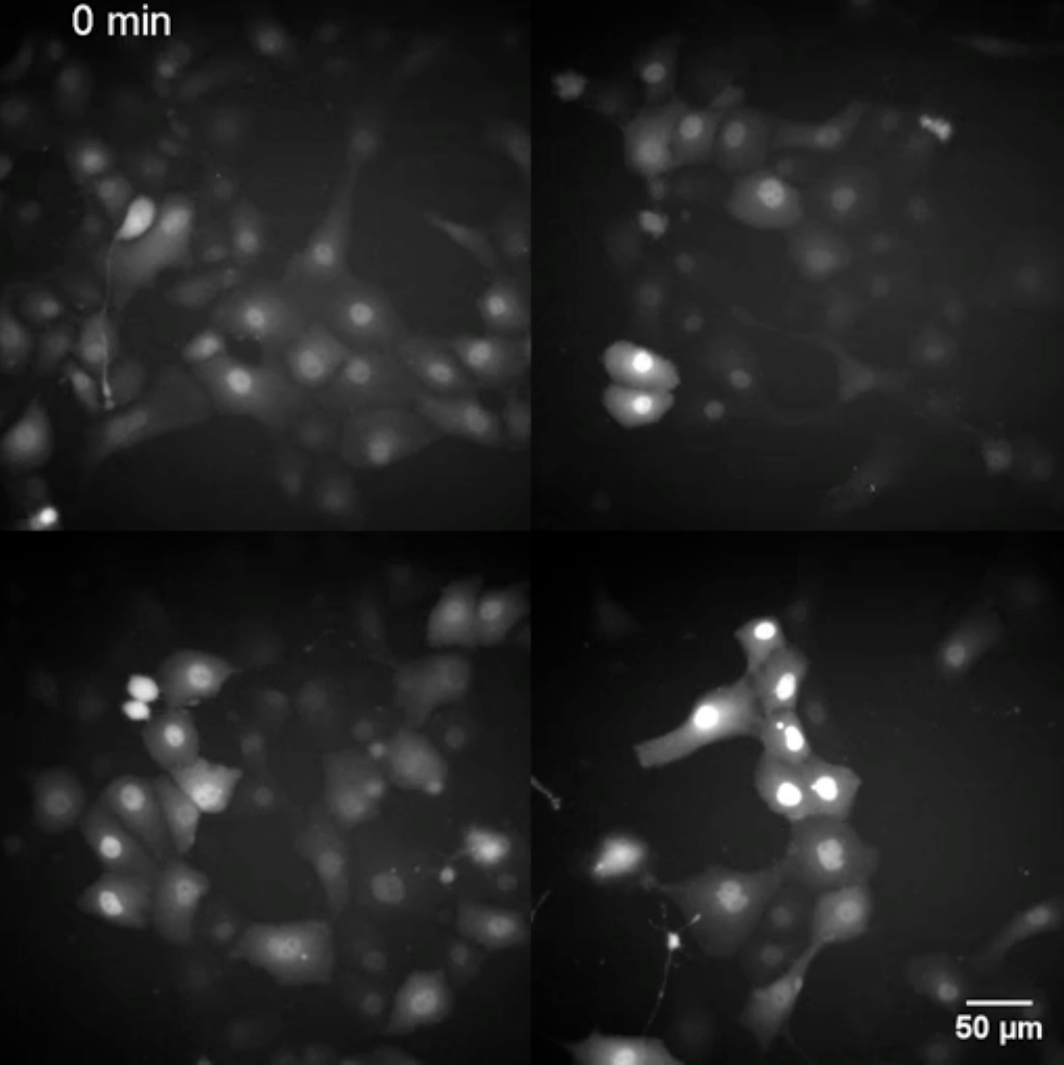}
    \caption{\textbf{Timelapse of the imaging, first main dataset.} Raw fluorescence microscopy images of the ERK-KTR-mKate2 reporter in cells of the dataset with 20 repeats. The four stage positions are here stitched side by side for easier visualization. An image is taken every \SI{2}{min}. The application of light stimuli starts at \SI{120}{min} and ends at \SI{4290}{min}.}
    \label{fig:SI-Movie1}
\end{video}

\begin{video}[h!]
    \centering
    \includegraphics[width=0.5 \linewidth]{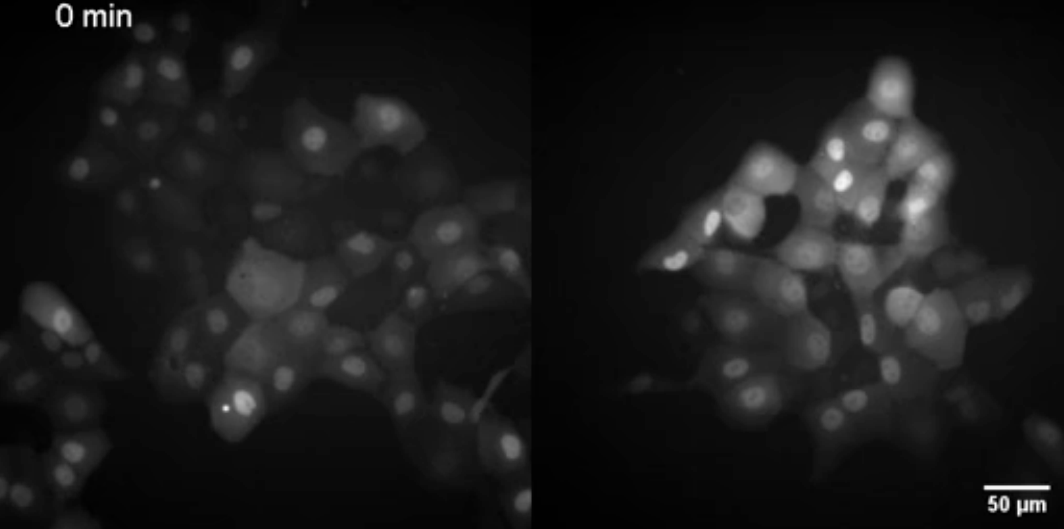}
    \caption{\textbf{Timelapse of the imaging, second main dataset.} Raw fluorescence microscopy images of the ERK-KTR-mKate2 reporter in cells of the dataset with 15 repeats. The two stage positions are stitched side by side for easier visualization. An image is taken every \SI{2}{min}. The application of light stimuli starts at \SI{100}{min} and ends at \SI{3220}{min}.}
    \label{fig:SI-Movie2}
\end{video}

\FloatBarrier

\bibliography{bibliography}